\pdfoutput=1
\documentclass[12pt,a4paper,reqno]{amsart}
\usepackage{amssymb}
\usepackage{amscd}
\usepackage[bookmarks,bookmarksnumbered,bookmarksopen,colorlinks,backref,linkcolor=blue,citecolor=red]{hyperref}%
\usepackage{enumerate}

\usepackage{graphicx}
\usepackage[rightcaption]{sidecap}
\usepackage{wrapfig}

\usepackage{siunitx}
\usepackage{tikz-cd}

\usepackage{color}
\usetikzlibrary{arrows}

\numberwithin{equation}{section}
\usepackage{mathabx}

\usepackage{cite}

\usepackage{mathtools}
\usepackage[tableposition=top]{caption}
\usepackage{booktabs,dcolumn}

\DeclareFontFamily{OT1}{rsfs}{}
\DeclareFontShape{OT1}{rsfs}{n}{it}{<-> rsfs10}{}
\DeclareMathAlphabet{\mathscr}{OT1}{rsfs}{n}{it}

\newcommand\M{\mathcal{M}}

\usepackage{algorithm, algorithmic}

\begin{document}

\title[A Comprehensive Review of Cauchy Stress Tensor, First Paper]{Cauchy Tetrahedron Argument and the Proofs of the Existence of Stress Tensor, a Comprehensive Review, Challenges, and Improvements}

\author[E. Azadi]{Ehsan Azadi${}^1$}
\footnotetext[1]{\Small{Department of Mechanical Engineering, Sharif University of Technology, Iran.\\ E-mail address: eazadi65@gmail.com}}
       

\maketitle

\begin{abstract}  
In 1822, Cauchy presented the idea of traction vector that contains both the normal and tangential components of the internal surface forces per unit area and gave the tetrahedron argument to prove the existence of stress tensor. These great achievements form the main part of the foundation of continuum mechanics. For about two centuries, some versions of tetrahedron argument and a few other proofs of the existence of stress tensor are presented in every text on continuum mechanics, fluid mechanics, and the relevant subjects. In this article, we show the birth, importance, and location of these Cauchy's achievements, then by presenting the formal tetrahedron argument in detail, for the first time, we extract some fundamental challenges. These conceptual challenges are related to the result of applying the conservation of linear momentum to any mass element, the order of magnitude of the surface and volume terms, the definition of traction vectors on the surfaces that pass through the same point, the approximate processes in the derivation of stress tensor, and some others. In a comprehensive review, we present the different tetrahedron arguments and the proofs of the existence of stress tensor, discuss the challenges in each one, and classify them in two general approaches. In the first approach that is followed in most texts, the traction vectors do not exactly define on the surfaces that pass through the same point, so most of the challenges hold. But in the second approach, the traction vectors are defined on the surfaces that pass exactly through the same point, therefore some of the relevant challenges are removed. We also study the improved works of Hamel and Backus, and indicate that the original work of Backus removes most of the challenges. This article shows that the foundation of continuum mechanics is not a finished subject and there are still some fundamental challenges.
\end{abstract}


\section{Introduction}
In 1822, for the first time, Cauchy in his lecture announced the forces on the surface of an internal mass element in continuum media in addition to the normal component on the surface can have the tangential components. An abstract of his lecture was published in 1823, \cite{Cau-1823}. In translation of Cauchy's lecture from the French by Maugin (2014, \cite{Maugin}), on page 50, we have:
\begin{quote}
\emph{However, the new ``pressure'' will not always be perpendicular to the faces on which it act, and is not the same in all directions at a given point.\\
\ldots Furthermore, the pressure or tension exerted on any plane can easily be deduced, in
both amplitude and direction, from the pressures or tensions exerted on three given
orthogonal planes. I had reached this point when M. Fresnel, who came to me to
talk about his works devoted to the study of light \ldots}
\end{quote}
Here the \emph{new pressure} is the traction vector that acts on the internal surface and contains both the normal and tangential components. Cauchy's works in continuum mechanics from 1822 to 1828 led to the derivation of \emph{Cauchy lemma} for traction vectors, the existence of \emph{stress tensor}, \emph{Cauchy equation of motion}, \emph{symmetry of stress tensor}, and some other achievements in the foundation of continuum mechanics \cite{Tr-Cla}. Cauchy's proof of the existence of stress tensor is called \emph{Cauchy tetrahedron argument}. From Truesdell (1971, \cite{Tr-Ther}), on page 8:
\begin{quote}
\emph{CAUCHY's theorem of the existence of the stress tensor, published in $1823$. CAUCHY, who knew full well the difference between a balance principle and a constitutive relation, stated the result clearly and proudly; he gave a splendid proof of it, which has been reproduced in every book on continuum mechanics from that day to this; and he recognized the theorem as being the foundation stone it still is.}
\end{quote}
On the importance of Cauchy's idea for traction vector and tetrahedron argument for the existence of stress tensor, Truesdell  (1968, \cite{Tr-Ess}), on page 188, says:
\begin{quote}
\emph{Clearly this work of Cauchy's marks one of the great turning points of mechanics and mathematical physics, even though few writers on the history of that subject seem to know it, a turning point that could well stand comparison with Huygens's theory of the pendulum, Newton's theory of the solar system, Euler's theory of the perfect fluid, and Maxwell's theories of the monatomic gas and the electromagnetic field.}
\end{quote}
This article gives a comprehensive review of the tetrahedron arguments and the proofs of the existence of stress tensor that represented during about two centuries, from 1822 until now, in many books and articles on continuum mechanics, fluid mechanics, solid mechanics, elasticity, strength of materials, etc. There are some different methods and processes to prove the existence of stress tensor and presentation of the Cauchy tetrahedron argument in the literature. We extract some fundamental challenges on these proofs and discuss these challenges in each one. To enter the subject, we first show the location of the Cauchy tetrahedron argument for the existence of stress tensor in the general steps of the foundation of continuum mechanics. Then, a formal proof of the Cauchy tetrahedron argument according to the accepted reference books will be given. We extract some fundamental challenges on this proof and discuss their importance in the foundation of continuum mechanics. Then we review different proofs in the literature and discuss their challenges. During this review, we also show the general approaches, important works, and their improvements.
\section{Location of Cauchy tetrahedron argument in the foundation of continuum mechanics}$\qquad$ \\
Although the birth of modern continuum mechanics is considered as the Cauchy's idea in 1822 \cite{Maugin}, some remarkable achievements were obtained earlier by famous mathematical physicians like Daniel Bernoulli, Euler, D'Alembert, Navier, Poisson, and the others. In general, these achievements can be addressed as the splitting of forces to the body forces and surface forces, the defining of pressure as the normal surface force per unit area, the considering of the internal mass element in continuum media, the Euler equation of motion, etc. But this was the genius of Cauchy to use the idea of his friend Fresnel, who worked on optics, in continuum mechanics and develop the idea of traction vector, the existence and properties of stress tensor, and the general equation of motion \cite{Cau-1823,Cau-1827,Maugin}. Cauchy's achievements were quickly taken as the foundation of continuum mechanics and the relevant subjects such as fluid mechanics, solid mechanics, elasticity, mechanics of deformable bodies, strength of materials, etc., \cite{Todhunter}. Recently, a good representation and description of the Cauchy's papers and the situation of continuum mechanics at that time was given by Maugin (2014, \cite{Maugin}).

The general steps that lead to the general concept of stress in continuum mechanics can be described as the following. Some of these steps were developed before Cauchy and others were developed or revised by Cauchy based on the new idea of traction vector that contains both the normal and tangential components on the surface.     
\begin{itemize}
\item The forces that apply to a fluid or solid element in continuum media can split to the \emph{surface forces} $(\boldsymbol{F}_s)$ and the \emph{body forces} $(\boldsymbol{F}_b)$, (before Cauchy). 
 \begin{equation}\label{Forces}
 \boldsymbol{F}=\boldsymbol{F}_s + \boldsymbol{F}_b
\end{equation}             
\item The surface force can be formulated as surface force per unit area that is called \emph{pressure} and is normal to the surface that it acts, (before Cauchy).\\
   
\item The surface force per unit area in addition to the normal component $(t_n)$ can have tangential components $(t_t)$. This general surface force per unit area is called \emph{traction vector}, (by Cauchy in 1822). 
 \begin{equation}\label{t_n-t_t}
 \boldsymbol{t}= t_n \boldsymbol{e}_n+t_t \boldsymbol{e}_t
 \end{equation} 
\item The traction vector depends only on the position vector $(\boldsymbol{r})$, time $(t)$, and the outward unit normal vector $(\boldsymbol{n})$ of the surface that acts on it in continuum media, (by Cauchy).
 \begin{equation}\label{t-r-t-n}
 \boldsymbol{t}=\boldsymbol{t}(\boldsymbol{r},t,\boldsymbol{n})
 \end{equation}
 \item The traction vectors acting on opposite sides of the same surface at a given point and time are equal in magnitude but opposite in direction. This is called \emph{Cauchy lemma}, (by Cauchy).
 \begin{equation}\label{t-n}
\boldsymbol{t}(\boldsymbol{r},t,\boldsymbol{n})= -\boldsymbol{t}(\boldsymbol{r},t,\boldsymbol{-n})
 \end{equation}
 \item 	\emph{Cauchy tetrahedron argument} states that the relation between the traction vector on a surface and the unit normal vector of that surface is linear, and this leads to the existence of a second order tensor that is called \emph{stress tensor}. The stress tensor $\boldsymbol{T}$ depends only on the position vector and time, (by Cauchy).
 \begin{equation}\label{t-tensor}
 \boldsymbol{t}= \boldsymbol{T}^T.\boldsymbol{n}
 \end{equation}
 where
\begin{equation}\label{tensor}
\boldsymbol{T}=\boldsymbol{T}(\boldsymbol{r},t)=
\begin{bmatrix}
T_{xx} & T_{xy} & T_{xz} \\
T_{yx} & T_{yy} & T_{yz} \\
T_{zx} & T_{zy} & T_{zz}
\end{bmatrix}
\end{equation}
\item Applying the conservation of linear momentum to a mass element in continuum media leads to the general differential equation of motion that is called \emph{Cauchy equation of motion}, (by Cauchy).
 \begin{equation}\label{Cau-eq}
 \rho \boldsymbol{a}= \nabla . \boldsymbol{T} + \rho \boldsymbol{b}
 \end{equation}
or
 \begin{equation}\label{Cau-eq-v}
 \rho (\frac{\partial\boldsymbol{v}}{\partial t}+(\boldsymbol{v}.\nabla)\boldsymbol{v})= \nabla . \boldsymbol{T} + \rho \boldsymbol{b}
 \end{equation}
where $\rho$, $\boldsymbol{b}$, $\boldsymbol{a}$, and $\boldsymbol{v}$ are the density, body force per unit mass, acceleration, and velocity, respectively.\\
\item The conservation of angular momentum shows that the stress tensor is symmetric, (by Cauchy).
 \begin{equation}\label{Txy-yx}
 T_{xy}=T_{yx}, \qquad T_{xz}=T_{zx}, \qquad T_{yz}=T_{zy}
 \end{equation}
 or
 \begin{equation}\label{T-T}
 \boldsymbol{T}=\boldsymbol{T}^T
 \end{equation}
\end{itemize}
These steps show the location of Cauchy tetrahedron argument for the existence of stress tensor in the foundation of continuum mechanics.
\section{Cauchy tetrahedron argument and the challenges}
The following representation of Cauchy tetrahedron argument is based on the two remarkable reference books on continuum mechanics, i.e., ``Truesdell and Toupin, The Classical Field Theories, pp. 542-543'' (1960, \cite{Tr-Cla}) and ``Malvern, Introduction to the Mechanics of a Continuous Medium, pp. 73-77'' (1969, \cite{Malvern}). Here we give more details to show clearly the process.
\subsection{Cauchy tetrahedron argument}
\begin{SCfigure}[20]
\includegraphics[width=8cm, clip, trim = 4.15cm 0.5cm 4.25cm 0.8cm]{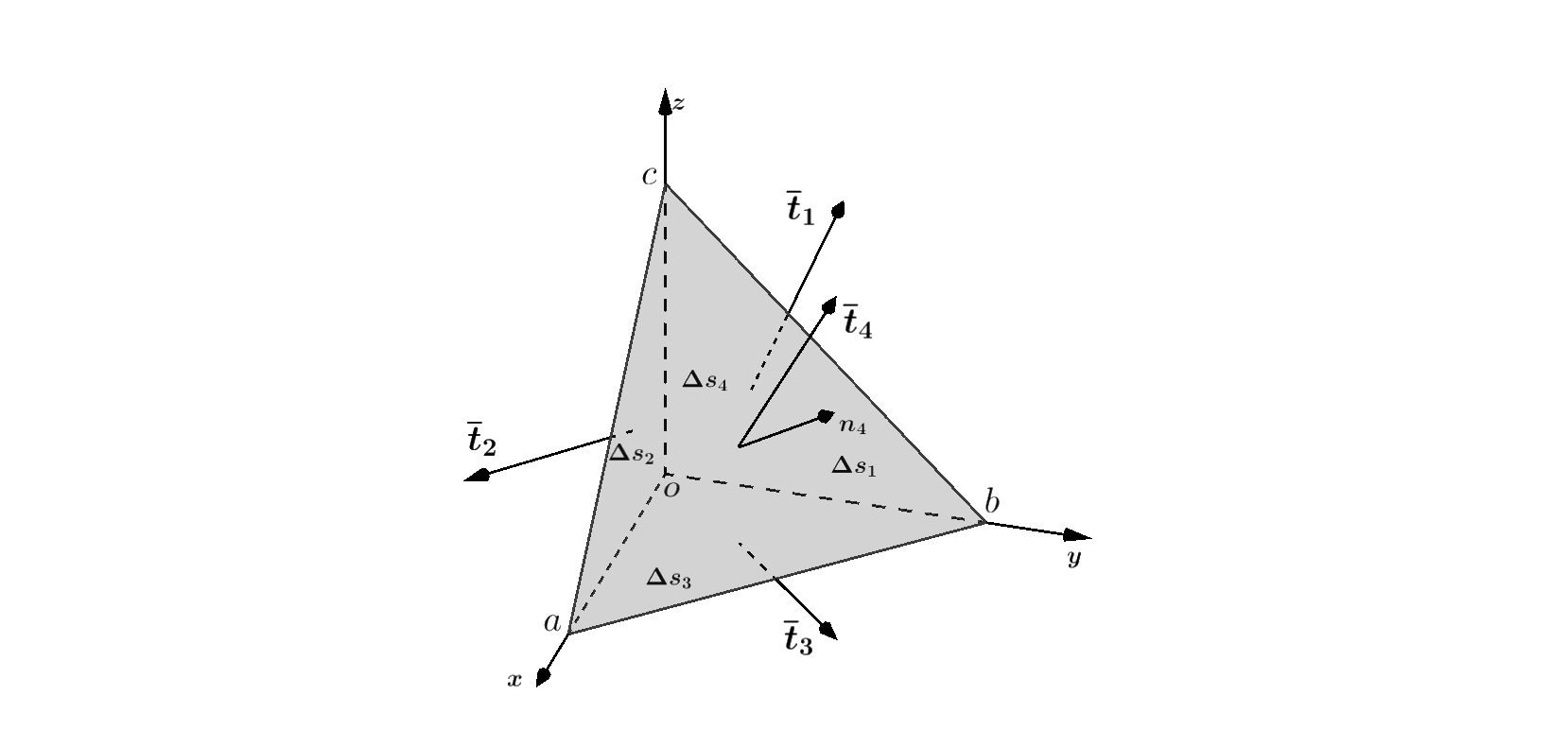}
\caption{\small{The geometry of tetrahedron element and the average traction vectors on the faces.}} 
\label{fig:tetrahedron}
\end{SCfigure}

$\qquad$ \\
Consider a tetrahedron element in continuum media that its vortex is at the point $\boldsymbol{o}$ and its three orthogonal faces are parallel to the three orthogonal planes of the Cartesian coordinate system. The fourth surface of the tetrahedron, i.e., its base, has the outward unit normal vector $\boldsymbol{n}_4$. The geometrical parameters and the average values of the traction vectors on the faces of tetrahedron are shown in Figure \ref{fig:tetrahedron}.
 The integral equation of conservation of linear momentum on a mass element $\M$ in continuum media is:
 \begin{equation}\label{Mom-eq}
 \int_{\partial \M}\boldsymbol{t}\, dS +  \int_{\M}\rho \boldsymbol{b}\, dV = \int_{\M}\rho \boldsymbol{a}\, dV
 \end{equation}
Now this law applies to the tetrahedron mass element. By averaging variables on the volume and faces of the tetrahedron element, the equation \eqref{Mom-eq} becomes:
 \begin{equation}\label{Mom-ave}
 \overline{\boldsymbol{t}}_4 \Delta s_4 +  \overline{\boldsymbol{t}}_1 \Delta s_1 +\overline{\boldsymbol{t}}_2 \Delta s_2 +\overline{\boldsymbol{t}}_3 \Delta s_3 +  \overline{\rho \boldsymbol{b}} \Delta V =   \overline{\rho \boldsymbol{a}} \Delta V   
 \end{equation}
 where the superscripts indicate the average values of these terms. The following geometrical relations for the faces and volume of tetrahedron hold:
  \begin{equation}\label{Ds_is}
 \begin{split}
  &\Delta s_1=n_x \Delta s_4, \qquad \Delta s_2=n_y \Delta s_4,  \qquad \Delta s_3=n_z \Delta s_4 \\ & 
 \Delta V= \frac{1}{3} h \Delta s_4   
  \end{split} 
 \end{equation}
 where $n_x$, $n_y$, and $n_z$ are components of the outward unit normal vector on $\Delta s_4$, i.e., $\boldsymbol{n}_4=n_x\boldsymbol{e}_x+n_y\boldsymbol{e}_y+n_z\boldsymbol{e}_z$. Here $h$ is the altitude of the tetrahedron. By substituting these geometrical relations into the equation \eqref{Mom-ave}:
 \begin{equation*}\label{Mom-ave-Ds-4}
 \overline{\boldsymbol{t}}_4 \Delta s_4 +  \overline{\boldsymbol{t}}_1 (n_x \Delta s_4) +\overline{\boldsymbol{t}}_2 (n_y \Delta s_4) +\overline{\boldsymbol{t}}_3 (n_z \Delta s_4) +  \overline{\rho \boldsymbol{b}} (\frac{1}{3} h \Delta s_4) =   \overline{\rho \boldsymbol{a}} (\frac{1}{3} h \Delta s_4)  
 \end{equation*} 
dividing through by $\Delta s_4$
 \begin{equation}\label{Mom-ave-h}
 \overline{\boldsymbol{t}}_4 + n_x \overline{\boldsymbol{t}}_1  + n_y \overline{\boldsymbol{t}}_2 +n_z \overline{\boldsymbol{t}}_3  +  \overline{\rho \boldsymbol{b}} (\frac{1}{3} h ) =   \overline{\rho \boldsymbol{a}} (\frac{1}{3} h)  
 \end{equation}
Now decrease the volume of tetrahedron element, $\Delta V\to 0$, in the way that $\boldsymbol{n}_4$ and the position of the vertex point of tetrahedron (point $\boldsymbol{o}$) do not change. As a result, $h\to 0$ and the tetrahedron shrinks to a point. So, in this limit, the body force and inertia term in the equation \eqref{Mom-ave-h} go to zero and the average traction vectors go to the exact values. The result is:
 \begin{equation}\label{t_4-t_is}
 \boldsymbol{t}_4 + n_x \boldsymbol{t}_1  + n_y \boldsymbol{t}_2 +n_z \boldsymbol{t}_3 = \boldsymbol{0}
 \end{equation}
 The traction vector $\boldsymbol{t}_1$ is applied to the surface $\Delta s_1$ by the unit normal vector $\boldsymbol{n}_1=-1\boldsymbol{e}_x$. Using the Cauchy lemma, i.e., $\boldsymbol{t}(\boldsymbol{r},t,\boldsymbol{n})=-\boldsymbol{t}(\boldsymbol{r},t,\boldsymbol{-n})$:
 \begin{equation}\label{t_n1}
\boldsymbol{t}(\boldsymbol{n}_1)=-\boldsymbol{t}(\boldsymbol{-n}_1)
 \end{equation}
 but $-\boldsymbol{n}_1=+1\boldsymbol{e}_x$ is the unit normal vector on the positive side of coordinate plane $yz$. If $\boldsymbol{t}_x$ is the traction vector on the positive side of coordinate plane $yz$, then by using the equation \eqref{t_n1}:
 \begin{equation}\label{t_n1x}
\boldsymbol{t}_1=-\boldsymbol{t}_x
 \end{equation}
 This strategy for $\boldsymbol{t}_2$ and $\boldsymbol{t}_3$ leads to:
\begin{equation}\label{t_n23}
\boldsymbol{t}_2=-\boldsymbol{t}_y , \qquad \boldsymbol{t}_3=-\boldsymbol{t}_z
 \end{equation}
By substituting these relations into the equation \eqref{t_4-t_is}: 
 \begin{equation*}\label{t_n4-txyz}
 \boldsymbol{t}_4 + n_x (-\boldsymbol{t}_x)  + n_y (-\boldsymbol{t}_y) +n_z (-\boldsymbol{t}_z) = \boldsymbol{0}
 \end{equation*}
 so
 \begin{equation}\label{t_n4-t_x_y_z}
 \boldsymbol{t}_4 = n_x \boldsymbol{t}_x  + n_y \boldsymbol{t}_y +n_z \boldsymbol{t}_z 
 \end{equation} 
 The traction vectors $\boldsymbol{t}_x$, $\boldsymbol{t}_y$, and $\boldsymbol{t}_z$ can be shown by their components as:
 \begin{equation}\label{t_compon}
 \begin{split}
 &\boldsymbol{t}_x = T_{xx} \boldsymbol{e}_x  + T_{xy} \boldsymbol{e}_y +T_{xz} \boldsymbol{e}_z \\ &
 \boldsymbol{t}_y = T_{yx} \boldsymbol{e}_x  + T_{yy} \boldsymbol{e}_y +T_{yz} \boldsymbol{e}_z \\ &
 \boldsymbol{t}_z = T_{zx} \boldsymbol{e}_x  + T_{zy} \boldsymbol{e}_y +T_{zz} \boldsymbol{e}_z
 \end{split}
 \end{equation} 
 By substituting these definitions into the equation \eqref{t_n4-t_x_y_z}:
 \begin{equation}\label{t_4-compon}
 \begin{split}
 \boldsymbol{t}_4 &= n_x(T_{xx} \boldsymbol{e}_x  + T_{xy} \boldsymbol{e}_y +T_{xz} \boldsymbol{e}_z) + n_y( T_{yx} \boldsymbol{e}_x  + T_{yy} \boldsymbol{e}_y +T_{yz} \boldsymbol{e}_z)\\ &{}
 + n_z(T_{zx} \boldsymbol{e}_x  + T_{zy} \boldsymbol{e}_y +T_{zz} \boldsymbol{e}_z)
 \end{split}
 \end{equation}
 or
  \begin{equation}\label{t_4-Txy}
 \begin{split}
 \boldsymbol{t}_4 &= (n_x T_{xx} + n_y T_{yx} +n_z T_{zx}) \boldsymbol{e}_x + (n_x T_{xy} + n_y T_{yy} +n_z T_{zy}) \boldsymbol{e}_y \\& 
 +(n_x T_{xz} + n_y T_{yz} +n_z T_{zz}) \boldsymbol{e}_z 
 \end{split}
 \end{equation}
 This can be shown as a relation between a second order tensor and a vector, as:
\begin{equation}\label{t_4-matrix}
\boldsymbol{t}_4=
\begin{bmatrix}
t_{x} \\
t_{y} \\
t_{z}
\end{bmatrix}_4=
\begin{bmatrix}
n_x T_{xx} + n_y T_{yx} +n_z T_{zx} \\
n_x T_{xy} + n_y T_{yy} +n_z T_{zy} \\
n_x T_{xz} + n_y T_{yz} +n_z T_{zz}
\end{bmatrix}=
\begin{bmatrix}
T_{xx} & T_{xy} & T_{xz} \\
T_{yx} & T_{yy} & T_{yz} \\
T_{zx} & T_{zy} & T_{zz}
\end{bmatrix}^T
\begin{bmatrix}
n_{x} \\
n_{y} \\
n_{z}
\end{bmatrix}_4
\end{equation}
therefore
\begin{equation}\label{t_4_tensor}
\boldsymbol{t}_4=\boldsymbol{T}^T .\boldsymbol{n}_4 
\end{equation}
By forming the tetrahedron element, no one of the components of $\boldsymbol{n}_4$ is zero. For the unit normal vectors that one or two of their components are equal to zero, the tetrahedron element does not form but due to the continuous property of the traction vectors on $\boldsymbol{n}$ and the arbitrary choosing for any orthogonal basis for the coordinate system, the equation \eqref{t_4_tensor} is valid for these cases, as well. So, the subscript $4$ can be removed from this equation:
\begin{equation}\label{tt-tensor}
\boldsymbol{t}=\boldsymbol{T}^T .\boldsymbol{n} 
\end{equation}
This equation shows that there is a second order tensor that is called stress tensor for describing the state of stress. This tensor $\boldsymbol{T}=\boldsymbol{T}(\boldsymbol{r},t)$, depends only on the position vector and time. Also, the relation between the traction vector on a surface and the unit normal vector of that surface is linear. 

Here the tetrahedron argument is finished. This argument and its result have a great importance and role in the foundation of continuum mechanics.

The following statements are not the elements of the tetrahedron argument and we state them to show the two other important achievements of Cauchy in the foundation of continuum mechanics. Cauchy applied the conservation of linear momentum to a ``cubic element'' and using his previous achievements, derived the general equation of motion that is called \emph{Cauchy equation of motion} \cite{Maugin}.
 \begin{equation}\label{motion}
 \rho \boldsymbol{a}= \nabla . \boldsymbol{T} + \rho \boldsymbol{b}
 \end{equation}
or
 \begin{equation}\label{v-motion}
 \rho (\frac{\partial\boldsymbol{v}}{\partial t}+(\boldsymbol{v}.\nabla)\boldsymbol{v})= \nabla . \boldsymbol{T} + \rho \boldsymbol{b}
 \end{equation}
 Also, by applying the conservation of angular momentum to a ``cubic element'', he showed that the stress tensor is symmetric \cite{Maugin}.
\begin{equation}\label{symmetry}
\boldsymbol{T}=\boldsymbol{T}^T
\end{equation}
\subsection{The challenges}$\qquad$\\
During study of the presented tetrahedron argument we found some conceptual challenges on it. In the following, we present and discuss them.
\begin{itemize}
\item Challenge 1: Note that applying the conservation of linear momentum to any mass element with any shape must lead to the general equation of motion that contains all of the effective terms including inertia, body forces, and surface forces (Cauchy equation of motion). But in this argument applying the conservation of linear momentum to the tetrahedron element leads to the equation \eqref{t_4-t_is}, i.e., $\boldsymbol{t}_4 + n_x \boldsymbol{t}_1  + n_y \boldsymbol{t}_2 +n_z \boldsymbol{t}_3 = \boldsymbol{0}$, that differs from the equation of motion \eqref{motion}, because the inertia and body forces do not exist in it. We saw that after presenting the tetrahedron argument, Cauchy and most of the authors derived the equation of motion by applying the conservation of linear momentum to a cubic element. What is the problem? applying the conservation of linear momentum to a tetrahedron element leads to the equation $\boldsymbol{t}_4 + n_x \boldsymbol{t}_1  + n_y \boldsymbol{t}_2 +n_z \boldsymbol{t}_3 = \boldsymbol{0}$ and the same process on a cubic element leads to the Cauchy equation of motion.\\

\item Challenge 2: The tetrahedron argument is based on the limit $\Delta V\to 0$, that is stated by all of the authors who presented this argument by the expressions like ``$\Delta V\to 0$'', ``$h\to 0$'', \emph{``when the tetrahedron shrinks to a point''}, or \emph{``when the tetrahedron shrinks to zero volume''}, while it must be proved that the existence of stress tensor at a point does not depend on the size of the considered mass element. In other words, the stress tensor exists for any size of mass element in continuum media where the volume of element increases, decreases, or does not change. By these proofs the result is valid only for the infinitesimal volumes and they did not show that this result can be applied to a mass element with any volume in continuum media.\\
 
\item Challenge 3: This tetrahedron argument is based on the average values of the effective terms in the integral equation of conservation of linear momentum and even for the limit $\Delta V\to 0$ this trend remains. While the stress tensor and the traction vectors relations are point-based and they must be derived from the exact point values.\\

 \item Challenge 4: During the tetrahedron argument we have the equation \eqref{Mom-ave-h}:
 \begin{equation*}\label{M-Av}
 \overline{\boldsymbol{t}}_4 + n_x \overline{\boldsymbol{t}}_1  + n_y \overline{\boldsymbol{t}}_2 +n_z \overline{\boldsymbol{t}}_3  +  \overline{\rho \boldsymbol{b}} (\frac{1}{3} h ) =   \overline{\rho \boldsymbol{a}} (\frac{1}{3} h)  
 \end{equation*}
 If we rewrite this equation as following, and take the limit:
  \begin{equation}\label{lim-mom}
\lim_{h \to 0}\Big(\frac{\overline{\boldsymbol{t}}_4 + n_x \overline{\boldsymbol{t}}_1  + n_y \overline{\boldsymbol{t}}_2 +n_z \overline{\boldsymbol{t}}_3}{\frac{1}{3}h}\Big)  = \lim_{h \to 0}(\overline{\rho \boldsymbol{a}}-\overline{\rho \boldsymbol{b}})
 \end{equation}
clearly the right hand side limit exists, because $\overline{\rho \boldsymbol{a}}-\overline{\rho \boldsymbol{b}}$ is bounded and is not generally equal to zero in continuum media. So, the left hand side limit must be existed and is not generally equal to zero. This implies that the order of magnitude of the denominator is $h$, i.e.:
\begin{equation}\label{O-t_is}
O(\overline{\boldsymbol{t}}_4 + n_x \overline{\boldsymbol{t}}_1  + n_y \overline{\boldsymbol{t}}_2 +n_z \overline{\boldsymbol{t}}_3) = h
 \end{equation}
so, $\overline{\boldsymbol{t}}_4 + n_x \overline{\boldsymbol{t}}_1  + n_y \overline{\boldsymbol{t}}_2 +n_z \overline{\boldsymbol{t}}_3$ and  $\overline{\rho \boldsymbol{a}} (\frac{1}{3} h ) -  \overline{\rho \boldsymbol{b}} (\frac{1}{3} h)$ have the same order of magnitude, that is, $h$. This means that by $h \to 0$ these two parts decrease by the same rate to zero and we cannot tell that the inertia and body terms go to zero faster than the surface terms. Since $O(\Delta s_4)=h^2$ and $O(\Delta V)=h^3$, we have:
 \begin{equation}\label{O-surface}
 \begin{split}
 &O(\Delta s_4(\boldsymbol{t}_4 + n_x \boldsymbol{t}_1  + n_y \boldsymbol{t}_2 +n_z \boldsymbol{t}_3)) \\
 &= O(\boldsymbol{t}_4 \Delta s_4+ \boldsymbol{t}_1 \Delta s_1+\boldsymbol{t}_2 \Delta s_2+\boldsymbol{t}_3 \Delta s_3) = h^3
 \end{split}    
 \end{equation} 
 and
  \begin{equation}\label{O-body}
O(\overline{\rho \boldsymbol{a}} \Delta V -  \overline{\rho \boldsymbol{b}}\Delta V)= h^3    
 \end{equation}       
so, we cannot tell that if $\Delta V \to 0$ or $h \to 0$ then the surface terms go to zero by $h^2$ and the inertia and body terms go to zero by $h^3$, because these two parts have the same order of magnitude, i.e., $h^3$, as shown above in \eqref{O-surface} and \eqref{O-body}.\\
 
\item Challenge 5: The purpose of Cauchy tetrahedron argument is to show that the traction vector at a point on a surface is a linear combination of the traction vectors on the three orthogonal surfaces that pass through that point. So, the four surfaces must pass through the same point to prove this relation between their traction vectors. But in the tetrahedron argument $\boldsymbol{t}_4$ is defined on the surface $\Delta s_4$ that does not pass through the vertex point of tetrahedron where the three surfaces $\Delta s_1$, $\Delta s_2$, and $\Delta s_3$ pass through it, see Figure \ref{fig:Inclined}.\\
   \begin{SCfigure}[20]
\includegraphics[width=7cm, clip, trim = 3.87cm 0.2cm 2.78cm 0.67cm]{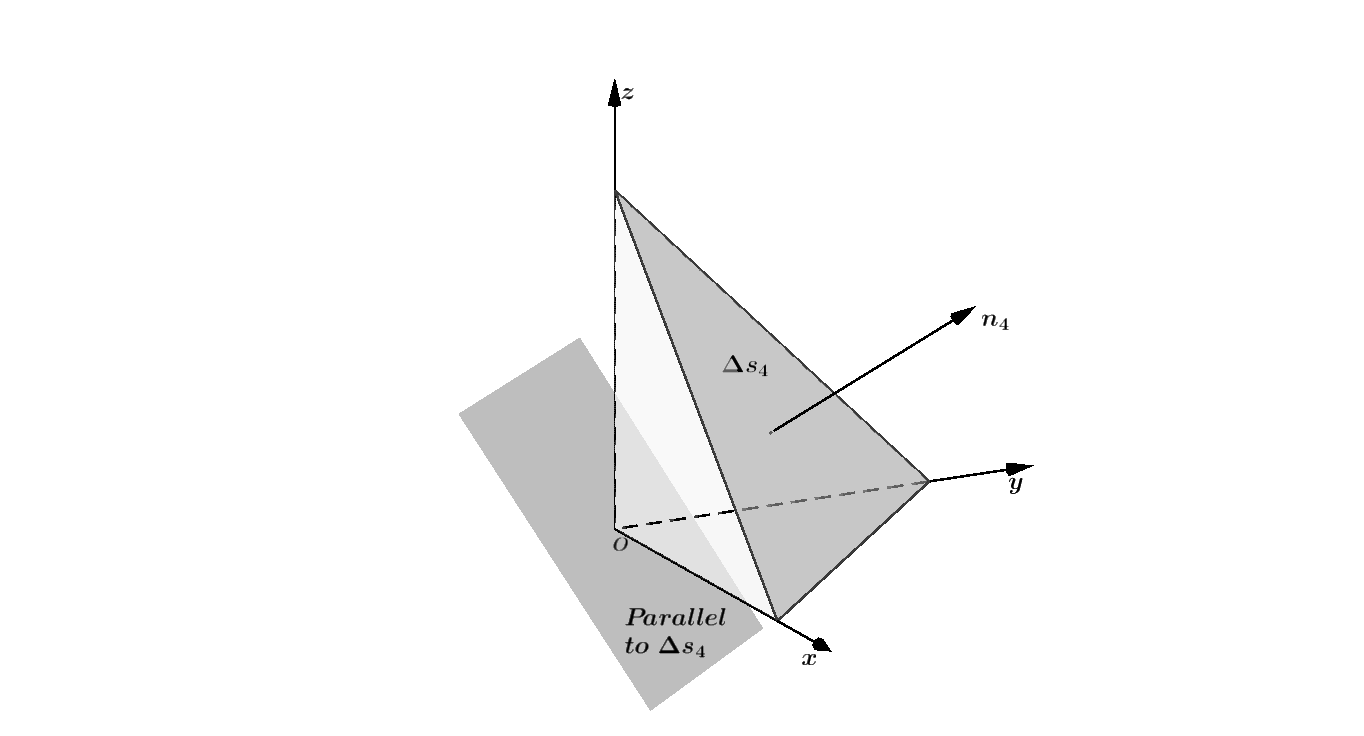}
\caption{\small{Inclined surface that is parallel to $\Delta s_4$ and passes through point $\boldsymbol{o}$.}}
\label{fig:Inclined}
 \end{SCfigure}
\item Challenge 6: The stress tensor is a point-based function. This means at any point in continuum media the stress tensor exists. So, in the equation $\boldsymbol{t}_4 + n_x \boldsymbol{t}_1  + n_y \boldsymbol{t}_2 +n_z \boldsymbol{t}_3 = \boldsymbol{0}$ the four traction vectors must belong to a unit point to conclude from the tetrahedron argument that $\boldsymbol{t}_4$ is related to a tensor that forms by the components of $\boldsymbol{t}_1$, $\boldsymbol{t}_2$, and $\boldsymbol{t}_3$. While in this proof, the surface that $\boldsymbol{t}_4$ is defined on it, i.e., $\Delta s_4$, does not pass through point $\boldsymbol{o}$, even for an infinitesimal tetrahedron element, see Figure \ref{fig:Inclined}.\\

\item Challenge 7: The result of this argument is the equation \eqref{t_4-t_is}, i.e., $\boldsymbol{t}_4 + n_x \boldsymbol{t}_1  + n_y \boldsymbol{t}_2 +n_z \boldsymbol{t}_3 = \boldsymbol{0}$, for an infinitesimal tetrahedron. Here the traction vectors are the average values on the faces of this infinitesimal tetrahedron. If we multiply this equation by $\Delta s_4$ that is the base area of the tetrahedron, the result is:
  \begin{equation}\label{Ds_4-t_iss}
 \Delta s_4(\boldsymbol{t}_4 + n_x \boldsymbol{t}_1  + n_y \boldsymbol{t}_2 +n_z \boldsymbol{t}_3) = \boldsymbol{t}_4 \Delta s_4+ \boldsymbol{t}_1 \Delta s_1+\boldsymbol{t}_2 \Delta s_2+\boldsymbol{t}_3 \Delta s_3 =\boldsymbol{0}
 \end{equation}
 but this is equal to the integral of $\boldsymbol{t}$ over the surface of $\M$, so:
 \begin{equation}\label{int_t_is}
 \boldsymbol{t}_4 \Delta s_4+ \boldsymbol{t}_1 \Delta s_1+\boldsymbol{t}_2 \Delta s_2+\boldsymbol{t}_3 \Delta s_3 = \int_{\partial \M}\boldsymbol{t}\, dS =\boldsymbol{0}  
 \end{equation}
where $\M$ is the infinitesimal tetrahedron element. This equation states that for the infinitesimal tetrahedron element the sum of the traction vectors on the surfaces of this element is zero. This means the surface forces have not any effect on the motion and acceleration of the element because their sum on the faces of element is zero. But this is not correct, since for any volume of mass element, even infinitesimal volume, the equation of conservation of linear momentum \eqref{Mom-eq}, the following equation, holds and tells us that this sum is not zero:
  \begin{equation*}\label{Momentom}
 \int_{\partial \M}\boldsymbol{t}\, dS +  \int_{\M}\rho \boldsymbol{b}\, dV = \int_{\M}\rho \boldsymbol{a}\, dV  
 \end{equation*}
 \item Challenge 8: In the previous challenge the equation \eqref{Ds_4-t_iss}, $\boldsymbol{t}_4 \Delta s_4+ \boldsymbol{t}_1 \Delta s_1+\boldsymbol{t}_2 \Delta s_2+\boldsymbol{t}_3 \Delta s_3 =\boldsymbol{0}$, states that the sum of surface forces on faces of the infinitesimal tetrahedron element is zero. So, it tells nothing about the relation between the traction vectors at a point on four different surfaces that pass through that point, because clearly $\boldsymbol{t}_4$ is defined on $\Delta s_4$ and this surface does not pass through point $\boldsymbol{o}$, even for an infinitesimal tetrahedron element, see Figure \ref{fig:Inclined}.
\end{itemize}
More discussions will be given in the next sections. 
\section{A comprehensive review}
The tetrahedron argument for the existence of stress tensor followed by many significant scientists and authors during about two centuries from 1822 to the present by some different versions. These proofs lead to the linear relation between the traction vector and the unit outward normal vector of the surface. This argument shows that the stress tensor exists and is independent of the surface characters. In the following, we show the different processes to prove this argument that exist in many textbooks on continuum mechanics and the relevant subjects such as fluid dynamics, solid mechanics, elasticity, plasticity, strength of materials, mathematical physics, etc.
\subsection{The first approach}$\qquad$ \\
Stokes in the famous article (1845, \cite{Stokes}), uses the Cauchy tetrahedron argument. On page 295:
\begin{quote}
\emph{\ldots Suppose now the dimensions of the tetrahedron infinitely diminished, then the resolved parts of the external and of the effective moving forces will vary ultimately as the cubes, and those of the pressures and tangential forces as the squares of homologous lines. \ldots \\
The method of determining the pressure on any plane from the pressures on three planes at right angles to each other, which has just been given, has already been employed by MM. Cauchy and Poisson.}
\end{quote}
So, from the part \emph{``now the dimensions of the tetrahedron infinitely diminished''} we can tell that Stokes's proof is based on infinitesimal volume. In this expression the inertia and body terms \emph{``vary ultimately as the cubes''} and surface terms vary \emph{``as the squares of homologous lines''}. While we showed in the challenge 4 that the surface terms and the inertia and body term vary by the same order of magnitude, that is, $h^3$.

Let see what is presented in the important book by Love, 1908. On pages 76-78 of the fourth edition of this book (1944, \cite{Love}), during the tetrahedron argument:
\begin{quote}
\emph{
 \begin{equation*}
 \iiint \rho f_x \, dxdydz = \iiint \rho X \, dxdydz + \iint X_v \, dS \qquad (1)
 \end{equation*}
[where $f_x$, $X$, and $X_v$ are acceleration, body force, and surface traction, respectively, all in the $x$ direction.]\footnote{The comments in the brackets [ ] are given by the author of the present article.}\\
46. Law of equilibrium of surface tractions on small volumes.\\
From the forms alone of equations (1) \ldots we can deduce a result of great importance. Let the volume of integration be very small in all its dimensions, and let $l^3$ denote this volume. If we divide both members of equation (1) by $l^2$, and then pass to a limit by diminishing $l$ indefinitely, we find the equation
 \begin{equation*}
 \lim_{l \to 0} l^{-2} \iint X_v \, dS =0
 \end{equation*}
 \ldots The equations of which these are types can be interpreted in the statement:\\ 
``The tractions on the elements of area of the surface of any portion of a body, which is very small in all its dimensions, are ultimately, to a first approximation, a system of forces in equilibrium.''\\
\ldots For a first approximation, when all the edges of the tetrahedron are small, we may take the resultant traction of the face [$\Delta s_4$]\ldots”} 
\end{quote}
So, here on these pages of Love's book, we see clearly the important challenges that are stated in the previous section. For example, \emph{``For a first approximation''}, \emph{``when all the edges of the tetrahedron are small''}, \emph{``Law of equilibrium of surface tractions on small volumes''}, \emph{``Let the volume of integration be very small''}, and clearly in the important statement inside the quotation marks that means for a first approximation, the summation of traction vectors on the surfaces of any portion of a body is zero when the portion is very small. We find that the Love's book is very important because it clearly and correctly represents the classical continuum mechanics in detail. For example, on these pages he has correctly stated that the results of Cauchy tetrahedron argument and the relation of traction vectors are approximate, for very small portion of body, and the relation between traction vectors is for the surfaces of mass element that do not pass through the same point. If instead of \emph{``divide both members of equation (1) by $l^2$''} we divide them by $l^3$, then the limit $l \to 0$ gives:
 \begin{equation*}
\lim_{l \to 0} l^{-3} \iint X_v \, dS = \lim_{l \to 0} l^{-3} \iiint \rho (f_x -X) \, dxdydz = \overline{\rho (f_x -X)}
 \end{equation*}
Similar to the challenge 4, here $\overline{\rho (f_x -X)}$ is a bounded value and is not generally equal to zero in continuum media. Therefore, for the existence of the limit on the left hand side the order of magnitude of the surface integral must be equal to $l^3$, i.e.:
\begin{equation*}
O\Big(\iint X_v \, dS \Big) = l^{3}
 \end{equation*}
that is equal to the order of magnitude of the volume integrals. So, the surface tractions are not in equilibrium even on small volumes, but are equal to the volume terms including inertia and body forces. By dividing \emph{``both members of equation $(1)$ by $l^2$''} the order of magnitude of these two parts is $l$, thus in the \emph{``limit by diminishing $l$ indefinitely''}, these two parts go to zero by the same rate. This is the trivial solution of the equation and cannot be a rigorous base for the existence of stress tensor. The proofs in some books are similar to the Love's proof, for example Planck (1932, \cite{Planck}), Serrin (1959, \cite{Serrin}), Aris (1989, \cite{Aris}), Marsden and Hughes (1994, \cite{Marsden}), Ogden (1997, \cite{Ogden}), Leal (2007, \cite{Leal}), Gonzalez and Stuart (2008, \cite{Gonzalez}). As a sample, in the book ``Vectors, Tensors, and the Basic Equations of Fluid Mechanics'' (1989, \cite{Aris}) by Aris, the proof on pages 100-101 is:
\begin{quote}
\emph{The principle of the conservation of linear momentum \ldots
 \begin{equation*}
 \frac{d}{dt}\iiint \rho \boldsymbol{v} \, dV = \iiint \rho \boldsymbol{f} \, dV + \iint \boldsymbol{t}_{(\boldsymbol{n})} \, dS \qquad (5.11.3)
 \end{equation*}
\ldots Suppose $V$ is a volume of given shape with characteristic dimension $d$. Then the volume of $V$ will be proportional to $d^3$ and the area of $S$ to $d^2$, with the proportionality constants depending only on the shape. Now let $V$ shrink on a point but preserve its shape, then the first two integrals in Eq. $(5.11.3)$ will decrease as $d^3$ but the last will be as $d^2$. It follows that
 \begin{equation*}
 \lim_{d \to 0} \frac{1}{d^2} \iint \boldsymbol{t}_{(\boldsymbol{n})} \, dS =\boldsymbol{0} \qquad (5.11.5)
 \end{equation*}
 or, the stresses are locally in equilibrium. \\
To elucidate the nature of the stress system at a point $P$ we consider a small tetrahedron with three of its faces parallel to the coordinate planes through $P$ and the fourth with normal $\boldsymbol{n}$. \ldots Then applying the principle of local equilibrium [Eq. $(5.11.5)$] to the 
stress forces when the tetrahedron is very small we have
 \begin{equation*}
 \begin{split}
 \boldsymbol{t}_{(\boldsymbol{n})} \, dA - \boldsymbol{t}_{(1)} \, dA_1 - \boldsymbol{t}_{(2)} \, dA_2 &{}- \boldsymbol{t}_{(3)} \, dA_3 \\&= (\boldsymbol{t}_{(\boldsymbol{n})} - \boldsymbol{t}_{(1)} \boldsymbol{n}_1 - \boldsymbol{t}_{(2)} \boldsymbol{n}_2 - \boldsymbol{t}_{(3)} \boldsymbol{n}_3)dA = \boldsymbol{0}.
 \end{split}
 \end{equation*}
 Now let $T_{ji}$ denote the $i^{th}$ component of $\boldsymbol{t}_{j}$ and $t_{(n)i}$ the $i^{th}$ component of $\boldsymbol{t}_{(\boldsymbol{n})}$ so that this equation can be written
  \begin{equation*}
 t_{(n)i} = T_{ji} n_j.
 \end{equation*}
}  
\end{quote}
Let us see what is presented for the existence of stress tensor in the Timoshenko's books. In the book ``Timoshenko and Goodier, Theory of Elasticity, 1934'', on page 213 according to the 1951 publication \cite{Timo-Theo}:
\begin{figure}[h]
\includegraphics[width=6cm]{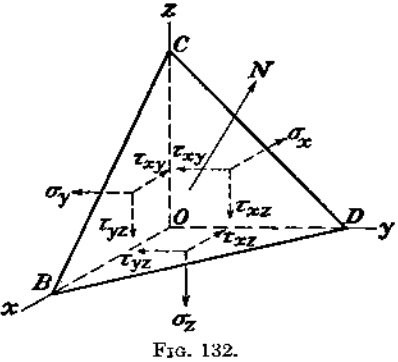}
\end{figure}
\begin{quote}
\emph{\ldots If these components of stress at any point are known, the stress acting on any inclined plane through this point can be calculated from the equations of statics [They considered only the case where acceleration is zero  and the body forces can be neglected, so there is no volume integral]. Let $O$ be a point of the stressed body and suppose the stresses are known for the coordinate planes \ldots (Fig. 132). To get the stress for any inclined plane through $O$, we take a plane $BCD$ parallel to it at a small distance from $O$, so that this latter plane together with the coordinate planes cuts out from the body a very small tetrahedron $BCDO$. Since the stresses vary continuously over the volume of the body, the stress acting on the plane $BCD$ will approach the stress on the parallel plane through $O$ as the element is made infinitesimal.\\
In considering the conditions of equilibrium of the elemental tetrahedron [acceleration is zero] the body forces can be neglected. Also as the element is very small we can neglect the variation of the stresses over the sides and assume that the stresses are uniformly distributed \ldots}
\end{quote}
Timoshenko has repeated almost the same process and comments in another book ``History of Strength of Materials'' (1953, \cite{Timo-His}). So, in these books we have the average values of the traction vectors on tetrahedron's faces and the traction vector on the base surface of the tetrahedron (surface $BCD$) is regarded as the traction vector on the inclined surface that is parallel to the surface $BCD$ and passes through point $O$. Therefore, most of the challenges hold. Also, this proof is limited to the cases that the mass element is in equilibrium (acceleration is zero) and the body forces are neglected. Similar process and assumptions are used for the tetrahedron argument by Prandtl and his coauthors (English translation 2004, \cite{Prandtl}). The proofs of the existence of stress tensor in some books are based on nearly similar process and assumptions to the above process, for example Sommerfeld (1950, \cite{Sommerfeld}), Biot (1965, \cite{Biot}), Feynman, Leighton, and Sands (1965, \cite{Feynman}) (using a wedge instead of a tetrahedron), Borg (1966, \cite{Borg}), Calcote (1968, \cite{Calcote}), Fl\"ugge (1972, \cite{Flugge}), Arfken (1985, \cite{Arfken}), Brekhovskikh and Goncharov (1994, \cite{Brekhovskikh}), Salencon (2001, \cite{Salencon}), Kundu, Cohen, and Dowling (2012, \cite{Kundu}), and Chaves (2013, \cite{Chaves}).

Let us see what is presented by Truesdell and Toupin in the very important book ``The Classical Field Theories, pp. 542-543'' (1960, \cite{Tr-Cla}):
\begin{quote}
\emph{\ldots Let the altitude of the tetrahedron be h;  the area of the inclined face [$\Delta s_4$],\ldots We may then estimate the volume integrals in $(200.1)$ [the integral equation of conservation of linear momentum] and apply the theorem of mean value to the surface integral:
 \begin{equation*}
\Delta s_4 (n_1 \boldsymbol{t}_1^{*}+n_2 \boldsymbol{t}_2^{*}+n_3 \boldsymbol{t}_3^{*}+\boldsymbol{t}_{(\boldsymbol{n})}^{*})+h \Delta s_4 K=\boldsymbol{0}, \qquad (203.1)
 \end{equation*}
where $K$ is a bound and where $\boldsymbol{t}_{(\boldsymbol{n})}^{*}$ [traction vector on $\Delta s_4$] and $\boldsymbol{t}_a^{*}$ [$\boldsymbol{t}_1^{*}$, $\boldsymbol{t}_2^{*}$, and $\boldsymbol{t}_3^{*}$] are the stress vectors at certain points upon the outsides of the respective faces. We cancel $\Delta s_4$ and let $h$ tend to zero, so obtaining 
  \begin{equation*}
\boldsymbol{t}_{(\boldsymbol{n})} = -(\boldsymbol{t}_1 n_1+\boldsymbol{t}_2 n_2+\boldsymbol{t}_3 n_3), \qquad (203.2)
 \end{equation*}
 where all stress vectors are evaluated at the vertex of the tetrahedron.}
\end{quote}
So, the expressions \emph{``then estimate the volume integrals''}, \emph{``apply the theorem of mean value to the surface integral''}, and \emph{``let $h$ tend to zero''} show the presented challenges in the before section. Here in the last line \emph{``where all stress vectors are evaluated at the vertex of the tetrahedron''} is not exactly obtained and is only an approximate result by this process, because $\boldsymbol{t}_{(\boldsymbol{n})}$ is defined on the base surface of the tetrahedron ($\Delta s_4$) and this surface does not pass exactly through the vertex of the tetrahedron even when $h$ tends to zero.   

In the book ``Introduction to the Mechanics of a Continuous Medium'' (1969, \cite{Malvern}) by Malvern, on pages 73-76:
\begin{quote}
\emph{\ldots Imagine \ldots a tetrahedron or triangular pyramid bound by parts of the three coordinate planes through O and a fourth plane $ABC$ not passing through $O$, \ldots \\
\ldots The asterisks indicate average values; thus $\boldsymbol{b}^*$ is the average value of the body force per unit mass in the tetrahedron. $\boldsymbol{t}^{(\boldsymbol{n})*}$ is the average value of the surface traction per unit area on the oblique face; \ldots \\
\ldots then the altitude $h$ will be allowed to approach zero so that the volume and the four surface areas simultaneously approach zero, while the orientation of $ON$ and the position of $O$ do not change. We postulate the continuity of all the components of the stress vectors and the body force and the density as functions of position; it follows that the average values will approach the local values at the point $O$, and the result will be an expression for the traction vector $\boldsymbol{t}^{(\boldsymbol{n})}$  at the point $O$ in the terms of the three special surface stress vectors $\boldsymbol{t}^{(k)}$ at $O$ \ldots
 \begin{equation*}
\boldsymbol{t}^{(\boldsymbol{n})*} \Delta S + \rho^* \boldsymbol{b}^* \Delta V - \boldsymbol{t}^{(1)*} \Delta S_1 - \boldsymbol{t}^{(2)*} \Delta S_2 - \boldsymbol{t}^{(3)*} \Delta S_3 = \rho^*  \Delta V \frac{d\boldsymbol{v}^*}{dt}.
 \end{equation*}
 \ldots dividing through by $\Delta S$, and rearranging terms we obtain
 \begin{equation*}
\boldsymbol{t}^{(\boldsymbol{n})*} + \frac{1}{3}h\rho^* \boldsymbol{b}^* = \boldsymbol{t}^{(1)*} n_1 +\boldsymbol{t}^{(2)*} n_2 + \boldsymbol{t}^{(3)*} n_3 + \frac{1}{3}h \rho^* \frac{d\boldsymbol{v}^*}{dt}.
 \end{equation*}
 We now let $h$ approach zero. The last term in each member then approaches zero, while the vectors in the other terms approach the vectors at the point $O$ as is indicated by dropping the asterisks. The result is in the limit
 \begin{equation*}
\boldsymbol{t}^{(\boldsymbol{n})} = \boldsymbol{t}^{(1)} n_1 +\boldsymbol{t}^{(2)} n_2 + \boldsymbol{t}^{(3)} n_3 =\boldsymbol{t}^{(k)}n_k. \qquad (3.2.7)
 \end{equation*}
This important equation permits us to determine the traction $\boldsymbol{t}^{(\boldsymbol{n})}$ at a point, acting on an arbitrary plane through the point, when we know the tractions on only three mutually perpendicular planes through the point. Note that this result was obtained without any assumption of equilibrium. It applies just as well in fluid dynamics as in solid mechanics.}
\end{quote}
This proof is similar to the presented tetrahedron argument for introducing the Cauchy tetrahedron argument in the previous section. So, all of the stated challenges hold in this proof. For example, \emph{``plane $ABC$ not passing through $O$''}, \emph{``asterisks indicate average values''}, \emph{``the average values will approach the local values at the point $O$''}, and \emph{``let $h$ approach zero''}. Note that the postulate in the last paragraph is not exact but as Love has been told \cite{Love}, is by a first approximation.

Tetrahedron arguments in many books are nearly similar to the presented proofs by Truesdell and Toupin (1960, \cite{Tr-Cla}) and Malvern (1969, \cite{Malvern}), for example Jaunzemis (1967, \cite{Jaunzemis}), Ilyushin and Lensky (1967, \cite{Ilyushin}), Rivlin (1969, \cite{Rivlin}), Wang (1979, \cite{Wang}), Eringen (1980, \cite{Eringen}), Narasimhan (1993, \cite{Narasimhan}), Chandrasekharaiah and Debnath (1994, \cite{Chandra}), Shames and Cozzarelli (1997, \cite{Shames}), Mase (1999, \cite{Mase}), Kiselev, Vorozhtsov, and Fomin (1999, \cite{Kiselev}), Batchelor (2000, \cite{Batchelor}), Basar and Weichert (2000, \cite{Basar}), Guyon, Hulin, Petit, and Mitescu (2001, \cite{Guyon}), Haupt (2002, \cite{Haupt}), Talpaert (2002, \cite{Talpaert}), Jog (2002, \cite{Jog}), Spencer (2004, \cite{Spencer}), Hutter and J\"ohnk (2004, \cite{Hutter}), Han-Chin (2005, \cite{Han}), Antman (2005, \cite{Antman}), Batra (2006, \cite{Batra}), Dill (2007, \cite{Dill}), Graebel (2007, \cite{Graebel}), Irgens (2008, \cite{Irgens}), Bonet and Wood (2008, \cite{Bonet}), Nair (2009, \cite{Nair}), Wegner and Haddow (2009, \cite{Wegner}), Lai, Rubin, and Krempl (2010, \cite{Lai}), Epstein (2010, \cite{Epstein}), Slawinski (2010, \cite{Slawinski}), Reddy (2010, \cite{Reddy}), Lautrup (2011, \cite{Lautrup}), Dimitrienko (2011, \cite{Dimitrienko}), Capaldi (2012, \cite{Capaldi}), Byskov (2013, \cite{Byskov}), Rudnicki (2015, \cite{Rudnicki}), and others.

In the book ``Introduction to the Mechanics of a Continuous Medium'' (1965, \cite{Sedov-In}) by Sedov, on pages 130-131:
\begin{quote}
\emph{\ldots Consider the volume $V$ as an infinitesimal tetrahedron \ldots with faces $MCB$, $MAB$, and $MAC$ perpendicular to the coordinate axes and with face $ABC$ arbitrarily determined by an externally directed unit normal vector \ldots The stresses on the areas with the normals $\ni_1$, $\ni_2$, $\ni_3$, and $\boldsymbol{n}$ are denoted by $\boldsymbol{p}^1$, $\boldsymbol{p}^2$, $\boldsymbol{p}^3$, and $\boldsymbol{p_n}$, respectively.\\
\ldots In fact, applying ($4.7$) [the integral equation of conservation of linear momentum] to the masses of the volume that are inside the infinitesimal tetrahedron $MABC$ at the instant in question, we obtain
 \begin{equation*}
 \begin{split}
&(\rho\boldsymbol{a}-\rho\boldsymbol{F}).\frac{1}{3}Sh \\
&=(-\boldsymbol{p}^1 S \cos (\widehat{\boldsymbol{n}\ni_1}) -\boldsymbol{p}^2 S \cos (\widehat{\boldsymbol{n}\ni_2})-\boldsymbol{p}^3 S \cos (\widehat{\boldsymbol{n}\ni_3})+\boldsymbol{p_n} .S )+S.O(h),
\end{split}
 \end{equation*}
 where $S$ is the area of the bounding surface $ABC$ [$\Delta s_4$], and $h$ is the infinitesimal height of the tetrahedron; $O(h)$, is a quantity which tends to zero for $h \to 0$. Approaching the limit, as $h \to 0$, we obtain
  \begin{equation*}
\boldsymbol{p_n}=\boldsymbol{p}^1 \cos (\widehat{\boldsymbol{n}\ni_1}) +\boldsymbol{p}^2 \cos (\widehat{\boldsymbol{n}\ni_2})+\boldsymbol{p}^3 \cos (\widehat{\boldsymbol{n}\ni_3}) \qquad (4.10)
 \end{equation*}
 }
\end{quote}
In this book, we see ``$O(h)$'' that represents the first order approximation. In addition, the \emph{``infinitesimal height of the tetrahedron''} and \emph{``tends to zero for $h \to 0$''} show that this proof, similar to earlier books, holds only for an infinitesimal tetrahedron. Nearly the same process is given in the other Sedov's book (1971, \cite{Sedov-Cour}).

In the book ``Theoretical Elasticity'' (1968, \cite{Green}) by Green and Zerna, on page 70:
\begin{quote}
\emph{
 \begin{equation*}
 \int_{\tau} \rho (\boldsymbol{F}_i-\dot{\boldsymbol{\omega}_i}) \, d\tau + \int_A \boldsymbol{t}_i \, dA =\boldsymbol{0}, \qquad (2.7.7)
 \end{equation*}
\ldots We consider a tetrahedron element bounded by the coordinate planes at the point $y_i$ and a plane whose unit normal is $\boldsymbol{n}_k$ measured from inside to outside of the tetrahedron. If we apply $(2.7.7)$ to this tetrahedron and take the limit as the tetrahedron tends to zero with $\boldsymbol{n}_k$ being unaltered we have
 \begin{equation*}
 \boldsymbol{t}_i =\boldsymbol{n}_k \boldsymbol{\sigma}_{ki}, \qquad (2.7.9)
 \end{equation*}
 Provided the contributions from the volume integrals may be neglected compared with the surface integrals, in the limit.}
\end{quote}
So, the challenges related to the \emph{``tetrahedron tends to zero''}, \emph{``volume integrals may be neglected compared with the surface integrals, in the limit''}, and definition of traction vectors on the surfaces that do not pass through the same point remain.

A more general proof is provided by Gurtin and his coauthors \cite{Gur-Mar,Gur-Mar2,Gur-An,Gur-2010}. Here it is represented from the book ``The Mechanics and Thermodynamics of Continua'' (2010, \cite{Gur-2010}) by Gurtin, Fried, and Anand. On pages 137-138:
\begin{quote}
\emph{A deep result central to all of continuum mechanics is \ldots Cauchy's theorem \ldots 
 \begin{equation*}
 \boldsymbol{t}(\boldsymbol{a},\boldsymbol{x})=-\sum_{i=1}^3 (\boldsymbol{a}.\boldsymbol{e}_i)\boldsymbol{t}(-\boldsymbol{e}_i,\boldsymbol{x}) \qquad (19.24)
 \end{equation*}
PROOF. Let $\boldsymbol{x}$ belong to the interior of $\mathcal{B}_t$. Choose $\delta>0$ and consider the (spatial) tetrahedron $\Gamma_{\delta}$ with the following properties: The faces of $\Gamma_{\delta}$ are $S_{\delta}$, $S_{1\delta}$, $S_{2\delta}$, and $S_{3\delta}$, where $\boldsymbol{a}$ and $-\boldsymbol{e}_i$ are the outward unit normals on $S_{\delta}$ and $S_{i\delta}$, respectively; the vertex opposite to $S_{\delta}$ is $\boldsymbol{x}$; the distance from $\boldsymbol{x}$ to $S_{\delta}$ is $\delta$. Then, $\Gamma_{\delta}$ is contained in the interior of $\mathcal{B}_t$ for all sufficiently small $\delta$, say $\delta \leq\delta_0$.\\
Next, if we assume that $\boldsymbol{b}$ [generalized body term including the inertia and body force] is continuous, then $\boldsymbol{b}$ is bounded on $\Gamma_{\delta}$. If we apply the force balance ($19.16$) to the material region $P$ occupying the region Γ$\Gamma_{\delta}$ in the deformed region at time $t$, we are then led to the estimate
 \begin{equation*}
 \Big|\int_{\partial\Gamma_{\delta}} \boldsymbol{t}(\boldsymbol{n}) \, da \Big|= \Big|\int_{\Gamma_{\delta}} \boldsymbol{b} \, dv \Big| \leq k \, vol(\Gamma_{\delta}) \qquad (19.25)
 \end{equation*}
for all $\delta \leq\delta_0$, where $k$ is independent of $\delta$.\\
Let $A(\delta)$ denote the area of $S_{\delta}$. Since $A(\delta)$) is proportional to $\delta^2$, while $vol(\Gamma_{\delta})$ is proportional to $\delta^3$, we may conclude from ($19.25$) that
 \begin{equation*}
\frac{1}{A(\delta)} \int_{\partial\Gamma_{\delta}} \boldsymbol{t}(\boldsymbol{n}) \, da \to \boldsymbol{0}
 \end{equation*}
 as $\delta \to 0$. But
 \begin{equation*}
 \int_{\partial\Gamma_{\delta}} \boldsymbol{t}(\boldsymbol{n}) \, da = \int_{S_{\delta}} \boldsymbol{t}(\boldsymbol{a}) \, da + \sum_{i=1}^3 \int_{S_{i\delta}} \boldsymbol{t}(-\boldsymbol{e}) \, da
 \end{equation*}
and, assuming that $\boldsymbol{t}(\boldsymbol{n},\boldsymbol{x})$ is continuous in $\boldsymbol{x}$ for each $\boldsymbol{n}$, since the area of $S_{i\delta}$ is $A(\delta)(\boldsymbol{a}.\boldsymbol{e}_i)$,
 \begin{equation*}
\frac{1}{A(\delta)} \int_{\partial S_{\delta}} \boldsymbol{t}(\boldsymbol{a}) \, da \to \boldsymbol{t}(\boldsymbol{a},\boldsymbol{x})
 \end{equation*}
 and
 \begin{equation*}
\frac{1}{A(\delta)} \int_{\partial S_{i\delta}} \boldsymbol{t}(-\boldsymbol{e}_i) \, da \to (\boldsymbol{a}.\boldsymbol{e}_i)\boldsymbol{t}(-\boldsymbol{e}_i,\boldsymbol{x}).
 \end{equation*} 
 Combining the relations above we conclude that $(19.24)$ is satisfied.
}
\end{quote}
This proof is based on the infinitesimal volume, and in the limit $\delta \to 0$ the traction vector on the base surface of tetrahedron is regarded as the traction vector on the inclined surface that passes through the vertex point of tetrahedron. The process that leads to
 \begin{equation*}
\frac{1}{A(\delta)} \int_{\partial\Gamma_{\delta}} \boldsymbol{t}(\boldsymbol{n}) \, da \to \boldsymbol{0}
 \end{equation*}
is similar to the Love's proof that was discussed before in detail. Therefore, some of the challenges remain. The proofs in some books are nearly similar to this process, for example Ciarlet (1988, \cite{Ciarlet}), Smith (1993, \cite{Smith}), Huilgol and Phan-Thien (1997, \cite{Huilgol}), Atkin and Fox (2005, \cite{Atkin}), Oden (2011, \cite{Oden}), Bechtel and Lowe (2015, \cite{Bechtel}).   

There is a new proof in the literature that is introduced by this statement: \emph{``This proof was furnished by W. Noll (private communication) in 1967.''}, in the chapter ``The Linear Theory of Elasticity'' by Gurtin in the book \cite{Tr-Gur}. Then this proof was presented in the two other books by Truesdell (1997, \cite{Tr-Gen}) and Liu (2002, \cite{Liu}). This proof is based on the properties of a linear transformation on vector space \cite{Noll-Light}. In the book by Leigh (1968, \cite{Leigh}), it is stated that if a transformation such as $\boldsymbol{T}$ on a vector space has the following properties then it is usually called a \emph{``linear transformation or tensor''}. On page 28 of this book \cite{Leigh}:
\begin{quote}
\emph{\ldots linear transformation $\boldsymbol{T}$ \ldots is defined by 
\begin{equation*}
 \begin{split}
&(a) \quad \boldsymbol{T}(\boldsymbol{u}+\boldsymbol{v})= \boldsymbol{T}(\boldsymbol{u})+\boldsymbol{T}(\boldsymbol{v})\\
&(b) \quad \boldsymbol{T}(\alpha\boldsymbol{v})= \alpha \boldsymbol{T}(\boldsymbol{v}) \qquad \qquad \qquad (2.8.1)
 \end{split} 
 \end{equation*}   
}
\end{quote}
Thus in the Noll's proof, it is tried to prove these properties for the traction vectors. These properties must be derived using the integral equation of conservation of linear momentum. This proof is nearly the same in the three books that are presented it \cite{Tr-Gur,Tr-Gen,Liu}. Here we represent it from the first book \cite{Tr-Gur}. On pages 48-49:
\begin{quote}
\emph{\ldots for any $\boldsymbol{x} \in B$ we can extend the function $\boldsymbol{s}(\boldsymbol{x},\boldsymbol{.})$ to all of $V$ as follows:
 \begin{equation*}
 \begin{split}
 &\boldsymbol{s}(\boldsymbol{x},\boldsymbol{v})=|\boldsymbol{v}|\boldsymbol{s}(\boldsymbol{x},\frac{\boldsymbol{v}}{|\boldsymbol{v}|}) \quad \boldsymbol{x} \neq \boldsymbol{0},\\
&\boldsymbol{s}(\boldsymbol{x},\boldsymbol{0})=\boldsymbol{0}.  \qquad  \qquad  \qquad \qquad \qquad (a)
  \end{split}
 \end{equation*}
 Let $\alpha$ be a scalar. If $\alpha > 0$, then
  \begin{equation*}
 \boldsymbol{s}(\alpha\boldsymbol{v})=|\alpha\boldsymbol{v}|\boldsymbol{s}\Big(\frac{\alpha\boldsymbol{v}}{|\alpha\boldsymbol{v}|}\Big)= \alpha|\boldsymbol{v}|\boldsymbol{s}\Big(\frac{\boldsymbol{v}}{|\boldsymbol{v}|}\Big)= \alpha \boldsymbol{s}(\boldsymbol{v}), \qquad (b)
 \end{equation*}
 where we have omitted the argument $\boldsymbol{x}$. If $\alpha < 0$, then $(b)$ and Cauchy's reciprocal theorem $(2)$ [$\boldsymbol{s}(\boldsymbol{n})= -\boldsymbol{s}(-\boldsymbol{n})$] yield
   \begin{equation*}
 \boldsymbol{s}(\alpha\boldsymbol{v})= \boldsymbol{s}(|\alpha|(-\boldsymbol{v}))= |\alpha|\boldsymbol{s}(-\boldsymbol{v})=\alpha\boldsymbol{s}(\boldsymbol{v}).
 \end{equation*}
 Thus $\boldsymbol{s}(\boldsymbol{x},\boldsymbol{.})$ is homogeneous.\\
To show that $\boldsymbol{s}(\boldsymbol{x},\boldsymbol{.})$ is additive we first note that
   \begin{equation*}
 \boldsymbol{s}(\boldsymbol{x},\boldsymbol{w}_1+\boldsymbol{w}_2)= \boldsymbol{s}(\boldsymbol{x},\boldsymbol{w}_1) + \boldsymbol{s}(\boldsymbol{x},\boldsymbol{w}_2)
 \end{equation*}
whenever $\boldsymbol{w}_1$ and $\boldsymbol{w}_2$ are linearly dependent. Suppose then that $\boldsymbol{w}_1$ and $\boldsymbol{w}_2$ are linearly independent. Fix $\epsilon > 0$ and consider $\pi_1$, the plane through $\boldsymbol{x}_0$ with normal $\boldsymbol{w}_1$; $\pi_2$, the plane through $\boldsymbol{x}_0$ with normal $\boldsymbol{w}_2$; and $\pi_3$, the plane through $\boldsymbol{x}_0+\epsilon\boldsymbol{w}_3$
with normal $\boldsymbol{w}_3$, where
   \begin{equation*}
 \boldsymbol{w}_3=-(\boldsymbol{w}_1+\boldsymbol{w}_2). \qquad (c)
 \end{equation*}
  \begin{figure}[htbt]
\includegraphics[width=7cm]{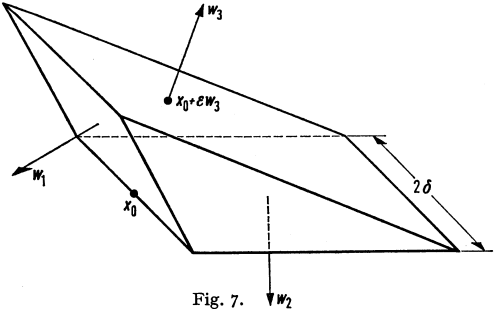}
\end{figure}
Consider the solid $\mathcal{A}=\mathcal{A}(\epsilon)$ bounded by these three planes and two planes parallel to both $\boldsymbol{w}_1$ and $\boldsymbol{w}_2$ and a distance $\delta$ from $\boldsymbol{x}_0$ (see Fig. $7$). Let $\epsilon$ and $\delta$ be sufficiently small that $\mathcal{A}$ is a part of $B$. Then
   \begin{equation*}
\partial\mathcal{A}=\bigcup^{5}_{i=1} \mathcal{W}_i,
 \end{equation*}
where $\mathcal{W}_i$, is contained in $\pi_i$ ($i =1,\, 2,\, 3$), and $\mathcal{W}_4$ and $\mathcal{W}_5$ are parallel faces. Moreover,
   \begin{equation*}
   \begin{split}
&a_i = \frac{|\boldsymbol{w}_i|}{|\boldsymbol{w}_3|}a_3  \qquad (i=1, \, 2), \\
&a_3 = O(\epsilon)  \qquad  as  \qquad \epsilon \to 0,\\
&\upsilon (\mathcal{A})=\frac{\epsilon}{2}|\boldsymbol{w}_3| a_3 = 2 \delta a_4=2 \delta a_5,
\end{split}
 \end{equation*}
 where $a_i$, is the area of $\mathcal{W}_i$. Thus, by the continuity of $\boldsymbol{s}_{\boldsymbol{n}}$,
 \begin{equation*}
\boldsymbol{c} \equiv \frac{|\boldsymbol{w}_3|}{a_3} \int_{\partial\mathcal{A}} \boldsymbol{s}_{\boldsymbol{n}} \, da = \sum^3_{i=1} \frac{|\boldsymbol{w}_i|}{a_i} \int_{\mathcal{W}_i} \boldsymbol{s}(\boldsymbol{x},\frac{\boldsymbol{w}_i}{|\boldsymbol{w}_i|}) \, da_{\boldsymbol{x}} + O(\epsilon) \quad as \quad \epsilon \to 0,
 \end{equation*}
and $(a)$ implies
 \begin{equation*}
\boldsymbol{c}= \sum^3_{i=1} \boldsymbol{s}(\boldsymbol{x}_0,\boldsymbol{w}_i)  + o(\boldsymbol{1}) \quad as \quad \epsilon \to 0.
 \end{equation*}
 On the other hand, we conclude from estimate $(a)$ [
 \begin{equation*}
|\int_{\partial P}\boldsymbol{s}_{\boldsymbol{n}} \, da |\leq k \upsilon (P)
 \end{equation*}
 where  $\upsilon (P)$ is the volume of $P$] in the proof of $(2)$ [$\boldsymbol{s}(\boldsymbol{n})= -\boldsymbol{s}(-\boldsymbol{n})$] that
 \begin{equation*}
\boldsymbol{c}=  O(\epsilon) \quad as \quad \epsilon \to 0.
 \end{equation*} 
 The last two results yield
  \begin{equation*}
\sum^3_{i=1} \boldsymbol{s}(\boldsymbol{x}_0,\boldsymbol{w}_i) =\boldsymbol{0}; 
 \end{equation*}
since $\boldsymbol{s}(\boldsymbol{x},\boldsymbol{.})$ is homogeneous, this relation and $(c)$ imply that $\boldsymbol{s}(\boldsymbol{x}_0,\boldsymbol{.})$ is additive. Thus $\boldsymbol{s}(\boldsymbol{x}_0,\boldsymbol{.})$ is linear, and, since $\boldsymbol{x}_0 \in B$ was arbitrarily chosen, Noll's proof is complete.}
\end{quote}
This is a creative proof by Noll that shows a new insight to the mathematical aspects of traction vectors. But this proof is based on the limited volume and holds for the infinitesimal mass element. The expression \emph{``$\epsilon \to 0$''} shows this. Also, the average values of the traction vectors are used and the traction vector on the surface $\pi_3$ is regarded as the traction vector on the surface that is parallel to $\pi_3$ and passes through point $\boldsymbol{x}_0$ in the limit. So, some of the challenges remain.

Leigh in the book ``Nonlinear Continuum Mechanics'' (1968, \cite{Leigh}) uses the properties of linear transformation to prove the existence of stress tensor by a different construction that is used in the Noll's proof. On pages 129-130:
\begin{quote}
\emph{
\begin{equation*}
\boldsymbol{t}=  \boldsymbol{f}(\boldsymbol{x},\boldsymbol{n}) \qquad \qquad (7.5.5)
 \end{equation*} 
  \begin{equation*}
\int_{\partial\chi} \boldsymbol{t} \, da + \int_{\chi} \boldsymbol{b} \rho \, dv  =  \int_{\chi} \boldsymbol{\ddot{x}} \rho \, dv \qquad (7.5.6)
 \end{equation*}
 Next we prove Cauchy's fundamental theorem for the stress
  \begin{equation*}
\boldsymbol{t}=  \boldsymbol{f}(\boldsymbol{x},\boldsymbol{n})=\boldsymbol{T}(\boldsymbol{x})\boldsymbol{n} \qquad (7.5.7)
 \end{equation*}
 \begin{figure}[h]
\includegraphics[width=8cm]{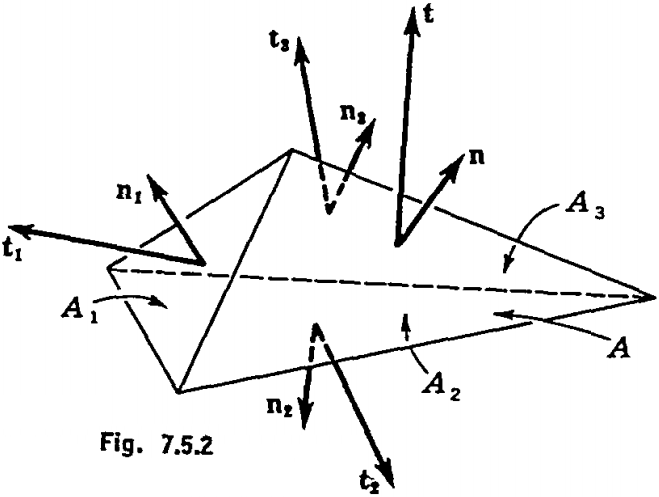}
\end{figure}
 that is, the stress vector $\boldsymbol{t}$ at $\boldsymbol{x}$ acting on the surface with direction $\boldsymbol{n}$ is a linear transformation of $\boldsymbol{n}$. The linear transformation or tensor $\boldsymbol{T}$ is called the stress tensor. Consider the elemental tetrahedron of Fig. $7.5.2$ \ldots The values of the stress vectors on the faces are given by $(7.5.5)$, where we use the same $\boldsymbol{x}$, since we are going to allow the tetrahedron to shrink to the point $\boldsymbol{x}$ in the limit. Thus we have
  \begin{equation*}
\boldsymbol{t}=  \boldsymbol{f}(\boldsymbol{x},\boldsymbol{n}) \qquad \qquad \qquad  \boldsymbol{t}_i=  \boldsymbol{f}(\boldsymbol{x},\boldsymbol{n}_i) \qquad (7.5.8)
 \end{equation*}   
 Thus applying $(7.5.6)$ to the elemental tetrahedron in the limit as $A,\, A_i \to 0$, we note that volume integrals are negligible compared with the surface integrals. The surface integral yields
  \begin{equation*}
\boldsymbol{t}= -\frac{1}{A} (A_1\boldsymbol{t}_1+A_2\boldsymbol{t}_2+A_3\boldsymbol{t}_3) \qquad (7.5.9)
 \end{equation*}
 Now a closed surface $S$ satisfies the condition
   \begin{equation*}
\int_{S} \boldsymbol{n} \, da = \boldsymbol{0} \qquad (7.5.10)
 \end{equation*}
 Applying $(7.5.10)$ to our elemental tetrahedron, we get
   \begin{equation*}
\boldsymbol{n}= -\frac{1}{A} (A_1\boldsymbol{n}_1+A_2\boldsymbol{n}_2+A_3\boldsymbol{n}_3) \qquad (7.5.11)
 \end{equation*} 
 Combining $(7.5.8)$, $(7.5.9)$, and $(7.5.11)$, we have, suppressing $\boldsymbol{x}$,
   \begin{equation*}
\boldsymbol{f}\Big(-\frac{1}{A} A_i\boldsymbol{n}_i \Big) = -\frac{1}{A} A_i\boldsymbol{f}(\boldsymbol{n}_i) \qquad (7.5.12)
 \end{equation*} 
 and we see that $\boldsymbol{f}(\boldsymbol{n})$ satisfies the definition $(2.8.1)$ of a linear transformation [two properties for linear transformation that we presented them before the Noll's proof], which proves $(7.5.7)$.}
\end{quote}
In this proof, Leigh uses three linearly independent traction vectors rather than two linearly traction vectors as used in the Noll's proof. As compared with the previous proofs that use a tetrahedron element with three orthogonal faces, in the Leigh's proof it is not needed the faces be orthogonal. But as previous proofs, this proof is based on the infinitesimal volume and is the sequence of the limit $A \to 0$. Here the \emph{``volume integrals are negligible compared with the surface integrals''} shows the challenge 4, so some of the challenges remain. The proof in the book by Lurie (2005, \cite{Lurie}) is similar to this proof.
\subsection{The second approach}$\qquad$ \\
During the comprehensive review of a large number of books on continuum mechanics and the relevant subjects, we found that there are two general approaches to the tetrahedron arguments and the proofs of the existence of stress tensor. In the first approach, the traction vectors and body terms are not defined at the same point. In fact, the traction vector on the base surface of the infinitesimal tetrahedron ($\Delta s_4$) is regarded as the traction vector on the inclined surface that is parallel to $\Delta s_4$ and passes through the vortex point of the tetrahedron. So, the challenges on the equation $\boldsymbol{t}_4+n_x\boldsymbol{t}_1+n_y\boldsymbol{t}_2+n_z\boldsymbol{t}_3=\boldsymbol{0}$ and most of the other stated challenges hold. Almost all the proofs in the previous subsection can be regarded in the first approach. Most of the tetrahedron arguments and the proofs of the existence of stress tensor are based on the first approach.

But in the second approach, the traction vectors and body terms are explicitly defined at the same point, e.g. in the tetrahedron arguments the vortex point ($\boldsymbol{o}$). Then by an approximate process for infinitesimal tetrahedron the equation $\boldsymbol{t}_{4_o}+n_x\boldsymbol{t}_{1_o}+n_y\boldsymbol{t}_{2_o}+n_z\boldsymbol{t}_{3_o}=\boldsymbol{0}$ is obtained. So, in the second approach all the traction vectors in this equation are exactly defined at the same point ($\boldsymbol{o}$) on different surfaces that pass exactly through this point. Here some of the challenges, for example challenges 6, 7, and 8 that are related to the definition of traction vectors at different points in the equation $\boldsymbol{t}_4+n_x\boldsymbol{t}_1+n_y\boldsymbol{t}_2+n_z\boldsymbol{t}_3=\boldsymbol{0}$ are removed. But these proofs are based on the approximate process and are limited to infinitesimal tetrahedron, so the other relevant challenges remain. A few of scientists and authors in continuum mechanics followed this approach. They are Muskhelishvili 1933 (English translation 1977, \cite{Muskh}), Sokolnikoff (1946, \cite{Soko}), Fung (1965, \cite{Fung-1965} and 1969, \cite{Fung-1969}), Godunov and Romenskii (1998, \cite{Godunov}), and Temam and Miranvilli (2000, \cite{Temam}). The proofs that are presented in all these books are nearly similar. Here, we present Muskhelishvili's proof and Fung's proof as two samples of these books. In the book ``Some Basic Problems of the Mathematical Theory of Elasticity, 1933'' by Muskhelishvili on pages 8-10 from the English translation, (1977, \cite{Muskh}):
\begin{quote}
\emph{Through the point $M$ draw three planes, parallel to the coordinate planes, and in addition, another plane having the normal $\boldsymbol{n}$ and lying a distance $h$ from $M$. These four planes form a tetrahedron, three faces of which are parallel to the coordinate planes, while the fourth $ABC$ [$\Delta s_4$] is the face to be considered. \ldots the transition to the limit $h\to 0$ the size of the tetrahedron will be assumed infinitely small.\newline
[here ($X_x, Y_x, Z_x$), ($X_y, Y_y, Z_y$), ($X_z, Y_z, Z_z$), and ($X_n, Y_n, Z_n$) are the components of traction vectors at the point $M$ on the four surfaces with unit normal vectors $\boldsymbol{e}_x$, $\boldsymbol{e}_y$, $\boldsymbol{e}_z$, and $\boldsymbol{n}$, respectively. $X$, $Y$, $Z$ are the components of body terms at the point $M$.]\newline
\ldots The projection of the body force equals $(X+\epsilon)dV$, where $dV$ is the volume of the tetrahedron. The value $X$ refers to the point $M$ and $\epsilon$ is an infinitely small quantity \ldots Further, the projection of the tractions, acting on the face $ABC$ is $(X_n+\epsilon')\sigma$ where $\sigma$ denotes the area of the triangle $ABC$ [$\Delta s_4$] and $\epsilon'$ is again infinitely small; $X_n$, $Y_n$, $Z_n$, as will be remembered, are the components of the stress vector acting on the plane through $M$ with normal $\boldsymbol{n}$.\\
Finally the projection of the external forces acting on $MBC$, normal to $Ox$, is $(- X_x + \epsilon_{1}) \sigma_{1}$ where $\sigma_1$ is the area of $MBC$. \ldots For the sides $MCA$ and $MAB$ one obtains similarly $(-X_y+\epsilon_2) \sigma_2$ and $(-X_z+\epsilon_3) \sigma_3$ respectively. Here $\epsilon_1$, $\epsilon_2$ and $\epsilon_3$ denote again infinitesimal quantities. [So, the conservation of linear momentum in $x$ direction is:]
\begin{equation*}
\begin{split}
&(X+\epsilon)\frac{1}{3}h\sigma + (X_n+\epsilon')\sigma + (- X_x + \epsilon_{1}) \sigma \cos (\boldsymbol{n},x) \\ &{}+(- X_y + \epsilon_{2}) \sigma \cos (\boldsymbol{n},y) + (- X_z + \epsilon_{3}) \sigma \cos (\boldsymbol{n},z)= 0.
\end{split}
 \end{equation*}
 Dividing by $\sigma$ and taking the limit $h\to 0$ one obtains the following formulae \ldots [similarly in $y$ and $z$ directions]:
 \begin{equation*}
 \begin{split}
&X_n = X_x \cos (\boldsymbol{n},x) + X_y \cos (\boldsymbol{n},y) + X_z \cos (\boldsymbol{n},z) \\
&Y_n = Y_x \cos (\boldsymbol{n},x) + Y_y \cos (\boldsymbol{n},y) + Y_z \cos (\boldsymbol{n},z) \\
&Z_n = Z_x \cos (\boldsymbol{n},x) + Z_y \cos (\boldsymbol{n},y) + Z_z \cos (\boldsymbol{n},z) 
\end{split}
\end{equation*}
}
\end{quote}
So, the traction vector on the inclined surface that passes exactly through point $M$ is obtained by an approximate process and by \emph{``taking the limit $h\to 0$''}. 

In the book ``A First Course in Continuum Mechanics, 1969'' by Fung, on pages 69-71 of the third edition, (1994, \cite{Fung-1969}):
\begin{quote}
\emph{Let us consider an infinitesimal tetrahedron formed by three surfaces parallel to the coordinate planes and one normal to the unit vector $\boldsymbol{v}$. Let the area of the surface normal to $\boldsymbol{v}$ be $dS$. \ldots \\
The forces in the positive direction of $x_1$, acting on the three coordinate surfaces, can be written as
 \begin{equation*}
(-\tau_{11} + \epsilon_1)dS_1, \qquad (-\tau_{21} + \epsilon_2)dS_2, \qquad (-\tau_{31} + \epsilon_3)dS_3,
 \end{equation*}
where $\tau_{11}$, $\tau_{21}$, $\tau_{31}$ are the stresses at the vertex $P$ opposite to $dS$. The negative sign 
is obtained because the outer normals to the three surfaces are opposite in sense with respect to the coordinate axes, and the $\epsilon$'s are inserted because the tractions act at points slightly different from $P$. If we assume that the stress field is continuous, then $\epsilon_1$, $\epsilon_2$, $\epsilon_3$ are infinitesimal quantities. On the other hand, the force acting on the triangle normal to $\boldsymbol{v}$ has a component $(T_1 + \epsilon)dS$ in the positive $x_1$-axis direction, the body force has an $x_1$-component equal to $(X_1 + \epsilon')dv$, and the rate of change of linear momentum has a component $\rho \dot{V}_1 dv$, where $\dot{V}_1$, is the component of acceleration in the direction of $x_1$. Here, $T_1$ and $X_1$ refer to the point $P$, and $\epsilon$ and $\epsilon'$ are again infinitesimal. The first equation of motion is thus
 \begin{equation*}
  \begin{split}
&(-\tau_{11} + \epsilon_1)v_1dS + (-\tau_{21} + \epsilon_2)v_2dS +(-\tau_{31} + \epsilon_3)v_3dS \\ &{}+ (T_1 + \epsilon)dS + (X_1 + \epsilon')\frac{1}{3}hdS = \rho \dot{V}_1 \frac{1}{3}hdS. \qquad \quad (3.3-3) 
 \end{split}
 \end{equation*}
Dividing through by $dS$, taking the limit as $h \to 0$, and noting that $\epsilon_1$, $\epsilon_2$, $\epsilon_3$, $\epsilon$, $\epsilon'$ vanish with $h$ and $dS$, one obtains
 \begin{equation*}
T_1=\tau_{11}v_1 + \tau_{21}v_2 + \tau_{31}v_3, \qquad (3.3-4)
 \end{equation*}   
}
\end{quote}
Fung has also discussed the error of this approximate process. On page 71, \cite{Fung-1969}:
\begin{quote}
\emph{{\small{\bfseries Checking Acceptable Errors}} \newline
\ldots We claimed that the sum of the terms
 \begin{equation*}
\epsilon_1v_1 + \epsilon_2v_2 + \epsilon_3v_3+ \epsilon + \frac{1}{3}(\epsilon'-\rho \dot{V}_1) \qquad (3.3-5)
 \end{equation*}
is small, compared with the terms that are retained; i.e., 
 \begin{equation*}
T_1, \, \tau_{11}v_1, \,\tau_{21}v_2, \, \tau_{31}v_3,  \qquad (3.3-6)
 \end{equation*}
when we take Eq. $(3.3-3)$ to the limit as $h \to 0$ and $\Delta S \to 0$. Now, if we are not allowed to take the limit as $h \to 0$ and $\Delta S \to 0$, but instead we are restricted to accept $h$ no smaller than a constant $h^*$ and $\Delta S$ no smaller than a constant multiplied by ($h^*)^2$, then the quantity listed in line $(3.3-5)$ must be evaluated for $h = h^*$ and $\Delta S = const_..(h^*)^2$ and must be compared with the quantities listed in line $(3.3-6)$. A standard of how small is negligible must be defined, and the comparison be made under that definition. If we find the quantity in line $(3.3-5)$ negligible compared with those listed in line $(3.3-6)$, then we can say that Eq. $(3.3-3)$ or Eq. $(3.3-2)$ [$T_i=v_j\tau_{ji}$] is valid. This tedious step should be done, in principle, to apply the continuum theory to objects of the real world.}
\end{quote}
\subsection{Advanced mathematical works}$\qquad$ \\
In recent decades, some proofs of the existence of Cauchy stress tensor or general Cauchy fluxes are presented in the geometrical language mathematics and advanced analysis. For example, using variational method \cite{Fosdick}, considering general Cauchy fluxes under weaker conditions \cite{Silhavy-84,Silhavy-90,Silhavy-91}, representing by measures \cite{Silhavy-08}, considering contact interactions as maps on pairs of subbodies and the possibility of handling singularities due to shocks and fracture \cite{Schuricht}, considering contact actions in N-th gradient generalized continua \cite{Isola}, etc. Each of these articles shows some aspects of the contact interactions in continuum physics. Here considering these attempts is outside the scope of this article that is based on the review of the proofs of the existence of stress tensor and their challenges in continuum mechanics and the relevant subjects.
\section{The work of Hamel, its improvements and challenges}
Let us see what is presented by Hamel in the famous book ``Theoretische Mechanik, pp. 513-514'' (1949, \cite{Hamel}). We present this proof completely:
\begin{quote}
\emph{Dann soll (I) nach Division mit dV die genauere Form
 \begin{equation*}
\varrho \overline{\boldsymbol{\omega}}=\sum \overline{\boldsymbol{\chi}} + \lim_{\Delta V \to 0} \frac{1}{\Delta V} \oint \boldsymbol{\overline{\sigma}}_{\boldsymbol{n}} \, dF  \qquad (I_A)
 \end{equation*}
bekommen und dieser Grenzwert existieren. Das Integral eistreckt sich \"uber die Oberfl\"ache des kleinen Volumens um den betrachteten Punkt, gegen den $\Delta V$ konvergiert.\\
Aus der Existenz des Grenzwertes folgen die S\"atze: 
\begin{equation*}
\boldsymbol{1})\qquad \overline{\boldsymbol{\sigma}}_n=\overline{\boldsymbol{\sigma}}_x \cos (\boldsymbol{n},x)+\overline{\boldsymbol{\sigma}}_y \cos (\boldsymbol{n},y)+\overline{\boldsymbol{\sigma}}_z \cos (\boldsymbol{n},z)
\end{equation*}
$\overline{\boldsymbol{\sigma}}_x$ usw. bedeuten die Spannungen an Fl\"achenelementen, deren \"au\ss ere Normalen Parallelen zur $x$, $y$, $z$-Achse sind. Setzt man
\begin{equation*}
\begin{split}
&\overline{\boldsymbol{\sigma}}_x = X_x \overline{\boldsymbol{i}} +Y_x \overline{\boldsymbol{j}}+Z_x \overline{\boldsymbol{k}}, \\
&\overline{\boldsymbol{\sigma}}_y = X_y \overline{\boldsymbol{i}} +Y_y \overline{\boldsymbol{j}}+Z_y \overline{\boldsymbol{k}}, \\
&\overline{\boldsymbol{\sigma}}_z = X_z \overline{\boldsymbol{i}} +Y_z \overline{\boldsymbol{j}}+Z_z \overline{\boldsymbol{k}}
\end{split}
\end{equation*}
mit $\overline{\boldsymbol{i}}$, $\overline{\boldsymbol{j}}$, $\overline{\boldsymbol{k}}$ als Einheitsvektoren in den drei Achsenrichtungen, so erscheint hier der Spannimgstensor
\begin{equation*}
\begin{Bmatrix}
X_x & Y_x & Z_x \\
X_y & Y_y & Z_y \\
X_z & Y_z & Z_z
\end{Bmatrix},
\end{equation*}
und man kann $\boldsymbol{1}$) auch schreiben:
\begin{equation*}
\boldsymbol{\overline{\sigma}}_{\boldsymbol{n}} = \overline{\overline{\sigma}} \, \overline{\boldsymbol{n}}
\end{equation*} 
wenn
\begin{equation*}
\overline{\boldsymbol{n}} = \overline{\boldsymbol{i}} \cos (\boldsymbol{n},x) + \overline{\boldsymbol{j}} \cos (\boldsymbol{n},y) + \overline{\boldsymbol{k}} \cos (\boldsymbol{n},z)
\end{equation*}  
den Einheitsvektor der \"au\ss eren Normalen angibt. (An der gedachten Fl\"ache wird also die Existenz einer solchen im allgemeinen vorausgesetzt.)\\
$\boldsymbol{1}$a) In $\boldsymbol{1}$) ist insbesondere enthalten
\begin{equation*}
\boldsymbol{\overline{\sigma}}_{-\boldsymbol{n}} = -\boldsymbol{\overline{\sigma}}_{\boldsymbol{n}},
\end{equation*}
d. h. das Gegenwirkungsprinzip f\"ur die inneren Spannungen, das also hier beweisbar ist.\\
$\boldsymbol{2}$) Die Ausf\"uhrung des Grenz\"uberganges in $(I_A)$ ergibt 
\begin{equation*}
\varrho \overline{\boldsymbol{\omega}}=\sum \overline{\boldsymbol{\chi}} + \frac{\partial \overline{\boldsymbol{\sigma}}_x}{\partial x} + \frac{\partial \overline{\boldsymbol{\sigma}}_y}{\partial y} + \frac{\partial \overline{\boldsymbol{\sigma}}_z}{\partial z}
\end{equation*}
}  
\end{quote}
Hamel's proof is based on the existence of the limit in the conservation of linear momentum equation $(I_A)$. This is the best part of this proof and the main improvement of his work. The original difference of this stage from the other previous similar works is that they divided the equation by $\Delta S$ that leads to the trivial solution. Because as indicated previously, the two parts (surface and volume integrals) of the equation have the same order of magnitude, i.e., $l^3$, and by dividing by $\Delta S$ they still have the same order of magnitude, i.e., $l$. So, they go to zero by the same rate when the element goes to infinitesimal volume, and this is a trivial result. But Hamel divided the equation by $\Delta V$ and this leads to the logical result of the existence the limit in the equation $(I_A)$. Therefore, some of the important challenges are removed by Hamel's proof.  

But this proof is limited to $\Delta V \to 0$ and there is no statement for a mass element with any volume size in continuum media. Because we must prove that the existence of stress tensor does not depend on the volume size of the considered mass element. So, the challenge 2 remains. In addition, there is no process to show how the equation $\overline{\boldsymbol{\sigma}}_n=\overline{\boldsymbol{\sigma}}_x \cos (\boldsymbol{n},x)+\overline{\boldsymbol{\sigma}}_y \cos (\boldsymbol{n},y)+\overline{\boldsymbol{\sigma}}_z \cos (\boldsymbol{n},z)$ is obtained from the existence of the limit in equation $(I_A)$. This will be an important step for the existence of stress tensor. 
\section{The work of Backus, its improvements and challenges}
Now let us see the Backus's proof from the book ``Continuum Mechanics'' (1997, \cite{Backus}). Unfortunately, the Backus's work seems to have attracted no attention of the scientists and authors in continuum mechanics, so far. However, this proof removes most of the challenges. First, we represent some notations according to this book. On page 163:
\begin{quote}
\emph{\ldots $(P, A_P)$ is oriented real physical space. \ldots The open set in $P$ occupied by the particles at time $t$ will be written $K(t)$, and the open subset of $K(t)$ consisting of the particles \ldots will be written $K'(t)$.}
\end{quote}
On pages 171-172:
\begin{quote}
\emph{\ldots $\vec{S}(\vec{r},t,\hat{n}_P)=\vec{S}_{force}(\vec{r},t,\hat{n}_P)+\vec{S}_{mfp}(\vec{r},t,\hat{n}_P)$ is called the stress on the surface $(S,\hat{n}_P)$. The total force exerted by the material just in front of 
$dA_P(\vec{r})$ on the material just behind $dA_P(\vec{r})$ is
 \begin{equation*}
d\vec{\mathcal{F}}_S (\vec{r})= dA_P(\vec{r})\vec{S}(\vec{r},t,\hat{n}_P). \qquad (13.2.7)
 \end{equation*}
 This is called the surface force on $dA_P(\vec{r})$.\ldots \\
\ldots Combining the physical law $(13.2.1)$ with the mathematical expressions $(13.2.3)$ and $(13.2.9)$ gives
 \begin{equation*}
\int_{K'} dV_P(\vec{r})(\rho \vec{a}-\vec{f})^E (\vec{r},t)= \int_{\partial K'} dA_P(\vec{r})\vec{S}(\vec{r},t,\hat{n}_P(\vec{r})). \qquad (13.2.10)
 \end{equation*}
 where $K'=K'(t)$ and $\hat{n}_P(\vec{r})$ is the unit outward normal to $\partial K'$ at $\vec{r} \in \partial K'$.}
 \end{quote}
In the following paragraphs, Backus has discussed some challenges. These are some aspects of the improvements of this work. On pages 172-173:
 \begin{quote}
\emph{To convert $(13.2.10)$ to a local equation, valid for all $\vec{r} \in K(t)$ at all times $t$, (i.e., to ``remove the integral signs'') we would like to invoke the vanishing integral theorem, \ldots The surface integral in $(13.2.10)$ prevents this.\\ 
Even worse, $(13.2.10)$ makes our model look mathematically self-contradictory, or internally inconsistent. Suppose that $K'$ shrinks to a point while preserving its shape. Let $\lambda$ be a typical linear dimension of $K'$. Then the left side of $(13.2.10)$ seems to go to zero like $\lambda^3$, while the right side goes to zero like $\lambda^2$. How can they be equal for all $\lambda >0?$ \\
Cauchy resolved the apparent contradiction in $1827$. He argued that the right side of $(13.2.10)$ can be expanded in a power series in $\lambda$, and the validity of $(13.2.10)$ for all $\lambda$ shows that the first term in this power series, the $\lambda^2$ term, must vanish. In modern language, Cauchy showed that this can happened iff at every instant $t$, at every $\vec{r} \in K(t)$, there is a unique tensor $\overleftrightarrow{S}^E(\vec{r},t)$ \ldots such that for each unit vector $\hat{n}$ \ldots,
 \begin{equation*}
\vec{S}(\vec{r},t,\hat{n})=\hat{n}.\overleftrightarrow{S}^E(\vec{r},t). \qquad (13.2.11)
 \end{equation*}
\ldots The physical quantity $\overleftrightarrow{S}$ is also called the Cauchy stress tensor.}
\end{quote}
Then on page 173 the Cauchy's theorem of the existence of stress tensor is stated:
 \begin{quote}
\emph{The argument which led Cauchy from $(13.2.10)$ to $(13.2.11)$ is fundamental to continuum mechanics, so we examine it in detail.\ldots \\
\newline
{\small{\bfseries Theorem $\mathbf{13.2.28}$ (Cauchy's Theorem)}} \ldots Suppose that for any open subset $K'$ of $K$ whose 
boundary $\partial K'$ is piecewise smooth, we have
 \begin{equation*}
\int_{K'} dV_U(\vec{r})\vec{f}(\vec{r})= \int_{\partial K'} dA_U(\vec{r})\vec{S}(\vec{r},\hat{n}_U(\vec{r})), \qquad (13.2.13)
 \end{equation*}
\ldots Then for each $\vec{r} \in K$ there is a unique $\overleftrightarrow{S}(\vec{r})$ \ldots such that for all $\hat{n}$ \ldots,
 \begin{equation*}
\vec{S}(\vec{r},\hat{n})=\hat{n}.\overleftrightarrow{S}(\vec{r}). \qquad (13.2.14)
 \end{equation*}
 }
 \end{quote}
Backus uses two lemmas to prove the ``Cauchy's Theorem''. The first lemma on pages 174-176:
\begin{figure}[htbh]
\includegraphics[width=8cm]{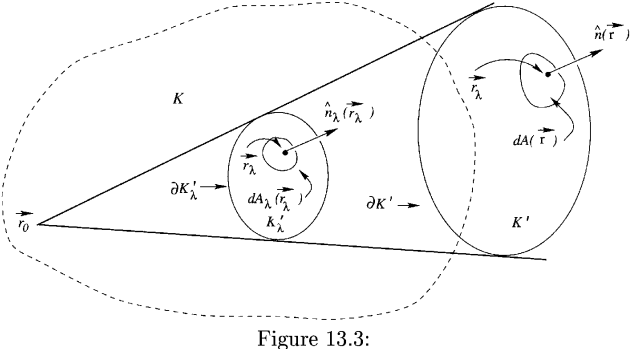}
\end{figure} 
\begin{quote}
\emph{Two lemmas are required. [The first lemma:]\\
{\small{\bfseries Lemma $\mathbf{13.2.29}$}} Suppose $\vec{f}$ and $\vec{S}$ satisfy the hypotheses of theorem $13.2.28$. Let $\vec{r}_0$ be any point in $K$ and let $K'$ be any open bounded (i.e., there is a real $M$ such that 
$\vec{r} \in K' \Rightarrow \|\vec{r}\| \leq M$) subset of $U$, with piecewise smooth boundary $\partial K'$. We don't need $K' \subseteq K$. Then
 \begin{equation*}
\int_{\partial K'} dA_U(\vec{r})\vec{S}(\vec{r}_0,\hat{n}_U(\vec{r}))=\vec{0}_V, \qquad (13.2.15)
 \end{equation*}
if $\hat{n}_U(\vec{r})$ is the unit outward normal to $\partial K'$ at $\vec{r} \in \partial K'$ and $dA_U(\vec{r})$ is the element of area on $\partial K'$.\\
\newline
{\small{\bfseries Proof of Lemma $\mathbf{13.2.29}$:}} For any real $\lambda$ in $0 < \lambda < 1$, define $\vec{\boldsymbol{r}}_{\lambda} : U \to U$ by $\vec{\boldsymbol{r}}_{\lambda} (\vec{r})= \vec{r}_0 + \lambda (\vec{r}-\vec{r}_0)$ for all $\vec{r} \in U$. Since $\vec{r}_{\lambda}(\vec{r}) - \vec{r}_0 = \lambda(\vec{r}-\vec{r}_0)$, $\vec{\boldsymbol{r}}_{\lambda}$ shrinks $U$ uniformly toward $\vec{r}_0$ by the factor $\lambda$. The diagram above is for $\lambda \approx 1/2$. Define $K'_{\lambda} = \vec{\boldsymbol{r}}_{\lambda}(K')$ so $\partial K'_{\lambda} = \vec{\boldsymbol{r}}_{\lambda}(\partial K')$. Choose $\vec{r} \in \partial K'$ and let $\vec{r}_{\lambda}= \vec{\boldsymbol{r}}_{\lambda}(\vec{r})$. Let $dA(\vec{r})$ denote a small nearly plane patch of surface in $\partial K'$, with $\vec{r} \in dA(\vec{r})$, and use  $dA(\vec{r})$ both as the name of this set and as the numerical value of its area. Let the set $dA_{\lambda}(\vec{r}_{\lambda})$ be defined as $\vec{\boldsymbol{r}}_{\lambda} (dA(\vec{r}))$, and denote its area also by $dA_{\lambda}(\vec{r}_{\lambda})$. Then by geometric similarity
 \begin{equation*}
dA_{\lambda}(\vec{r}_{\lambda}) = \lambda^2 dA(\vec{r}). \qquad (13.2.16)
 \end{equation*}
Let $\hat{n}(\vec{r})$ be the unit outward normal to $\partial K'$ at $\vec{r}$, and let $\hat{n}_{\lambda}(\vec{r}_{\lambda})$ be the unit outward normal to $\partial K'_{\lambda}$ at $\vec{r}_{\lambda}$. By similarity, $\hat{n}(\vec{r})$ and $\hat{n}_{\lambda}(\vec{r}_{\lambda})$ point in the same direction. Being unit vectors, they are equal:
 \begin{equation*}\label{tracion}
\hat{n}_{\lambda}(\vec{r}_{\lambda}) = \hat{n}(\vec{r}). \qquad (13.2.17)
 \end{equation*}
 Since $\vec{r}_0$ is fixed, it follows that
\begin{equation*}
\int_{\partial K'_{\lambda}} dA_{\lambda}(\vec{r}_{\lambda})\vec{S}(\vec{r}_0,\hat{n}_{\lambda}(\vec{r}_{\lambda})) = \lambda^2 \int_{\partial K'} dA(\vec{r})\vec{S}(\vec{r}_0,\hat{n}(\vec{r})) , \qquad (13.2.18)
 \end{equation*}
 If $\lambda$ is small enough, $K' \subseteq K$. Then, by hypothesis, we have $(13.2.13)$ with $K'$ and 
$\partial K'$ replaced by $K'_{\lambda}$ and $\partial K'_{\lambda}$. Therefore\footnote{In the first integral on the right hand side, $\partial$ is missed in the original book.}
 \begin{equation*}
  \begin{split}
\int_{K'_{\lambda}} dV(\vec{r})\vec{f}(\vec{r}) &= \int_{\partial K'_{\lambda}} dA_{\lambda}(\vec{r}_{\lambda}) \Big\{ \vec{S}(\vec{r}_{\lambda},\hat{n}_{\lambda}(\vec{r}_{\lambda})) - \vec{S}(\vec{r}_0,\hat{n}_{\lambda}(\vec{r}_{\lambda})) \Big\} \\ &{}+ \int_{\partial K'_{\lambda}} dA_{\lambda}(\vec{r}_{\lambda}) \vec{S}(\vec{r}_0,\hat{n}_{\lambda}(\vec{r}_{\lambda})).
 \end{split}
 \end{equation*}
 From $(13.2.18)$ it follows that
 \begin{equation*}
  \begin{split}
&\int_{\partial K'} dA(\vec{r})\vec{S}(\vec{r}_0,\hat{n}(\vec{r}))= \frac{1}{\lambda^2} \int_{K'_{\lambda}} dV(\vec{r})\vec{f}(\vec{r}) \\ &{}+ \frac{1}{\lambda^2} \int_{\partial K'_{\lambda}} dA_{\lambda}(\vec{r}_{\lambda}) \Big\{ \vec{S}(\vec{r}_{\lambda},\hat{n}_{\lambda}(\vec{r}_{\lambda})) - \vec{S}(\vec{r}_0,\hat{n}_{\lambda}(\vec{r}_{\lambda})) \Big\}. \qquad (13.2.19)
 \end{split}
 \end{equation*}
Let $m_{\vec{S}}(\lambda)= $ maximum value of $\|\vec{S}(\vec{r}_0, \hat{n})-\vec{S}(\vec{r}, \hat{n})\|$ for all $\vec{r} \in \partial K'_{\lambda}$ and all $\hat{n} \in N_U$.\\
Let $m_{\vec{f}}(\lambda)= $ maximum value of $|\vec{f}(\vec{r})|$ for all $\vec{r} \in K'_{\lambda}$. \\
Let $|\partial K'_{\lambda}|=$ area of $\partial K'_{\lambda}$, $|\partial K'|=$ area of $\partial K'$. \\
Let $|K'_{\lambda}|=$ volume of $K'_{\lambda}$, $|K'|=$ volume of $K'$.\\
Then $|\partial K'_{\lambda}|= \lambda^2 |\partial K'|$ and $|K'_{\lambda}|= \lambda^3 |K'|$, so $(10.2.3)$ and $(13.2.19)$ imply 
\begin{equation*}
\Big\| \int_{\partial K'} dA(\vec{r})\vec{S}(\vec{r}_0,\hat{n}(\vec{r})) \Big\| \leq \lambda |K'|m_{\vec{f}}(\lambda) + |\partial K'|m_{\vec{S}}(\lambda). \qquad (13.2.20)
 \end{equation*}
As $\lambda \to 0$, $m_{\vec{f}}(\lambda)$ remains bounded (in fact $\to \|f(\vec{r}_0)\|$) and $m_{\vec{S}}(\lambda)\to 0$ because $\vec{S}: K\times N_U \to V$ is continuous. Therefore, as $\lambda \to 0$, the right side of $(13.2.20)$ $\to 0$. Inequality $(13.2.20)$ is true for all sufficiently small $\lambda > 0$, and the left side is non-negative and independent of $\lambda$. Therefore the left side must be $0$. This proves $(13.2.15)$ and hence proves lemma $13.2.29$.}
\end{quote}
So, the result of this lemma is the fundamental equation $(13.2.15)$ for traction vectors at the given point $\vec{r}_0$, as the following:
 \begin{equation*}
\int_{\partial K'} dA_U(\vec{r})\vec{S}(\vec{r}_0,\hat{n}_U(\vec{r}))=\vec{0}_V
 \end{equation*}
If we compare this equation with the presented similar equations in the previous sections, the lemma $13.2.29$ and its proof are the improved achievements by Backus. Because:
\begin{itemize}
\item This equation is obtained by an exact process, not by an approximate process.
\item This equation is exactly valid not only for an infinitesimal volume where the volume of $K'$ tends to zero but also for any volume of $K'$ in continuum media.
\item In this integral equation the position vector is fixed at the point $\vec{r}_0$, so the stress vector changes only by changing the unit normal vector on the surface of the mass element at a given time. This is the key character that leads to the exact validation of this equation for any volume of mass element in continuum media. In the former proofs, stress vector changes by changing both the position vector $\vec{r}$ and the unit normal vector on surface of the mass element at a given time and this leads to the approximate proofs for only the mass elements with infinitesimal volumes.  
\end{itemize}
Backus uses a second lemma to prove the existence of stress tensor based on the equation $(13.2.15)$. On pages 176-180:   
\begin{quote}
\emph{We also need [The second lemma:]\\
{\small{\bfseries Lemma $\mathbf{13.2.30}$}} Suppose $\vec{S}: N_U \to V$. Suppose that for any open set $K'$ with piecewise 
smooth boundary $\partial K'$, $\vec{S}$ satisfies
 \begin{equation*}
\int_{\partial K'} dA(\vec{r})\vec{S}(\hat{n}(\vec{r}))=\vec{0}_V \qquad (13.2.21)
 \end{equation*}
 where $\hat{n}(\vec{r})$ is the unit outward normal to $\partial K'$ as $\vec{r}\in \partial K'$. Suppose $F : U \to V$ is defined as follows:
 \begin{equation*}
F(\vec{0}_U)=\vec{0}_V \,\,and \, \,if \,\, \vec{u} \neq \vec{0}_U, \, F(\vec{u})=\|\vec{u}\|\vec{S}(\frac{\vec{u}}{\|\vec{u}\|}). \qquad (13.2.22)
 \end{equation*}
 Then $F$ is linear.\\
 \newline 
 {\small{\bfseries Proof of Lemma $\mathbf{13.2.30}$:}} a) $F(—\vec{u}) = —F(\vec{u})$ for all $\vec{u} \in U$. To prove this, it suffices to prove
  \begin{equation*}
\vec{S}(-\hat{n})=-\vec{S}(\hat{n}) \, \, for \, \, all  \,  \, \hat{n} \in N_U. \qquad (13.2.23)
 \end{equation*}
 Let $K'$ be the flat rectangular box shown at upper right. For this box, $(13.2.21)$ 
gives 
  \begin{equation*}
L^2\vec{S}(\hat{n})+L^2\vec{S}(-\hat{n})+ \epsilon L (\vec{S}(\hat{n}_1)+\vec{S}(-\hat{n}_1)+\vec{S}(\hat{n}_2)+\vec{S}(-\hat{n}_2))=\vec{0}_V.
 \end{equation*}
 Hold $L$ fixed and let $\epsilon \to 0$. Then divide by $L^2$ and $(13.2.23)$ is the result.
 \begin{figure}[htbt]
\includegraphics[height=8cm]{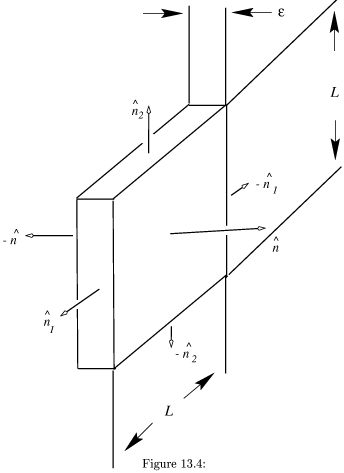}
\end{figure} \\
 b) If $c \in \mathcal{R}$ and $\vec{u} \in U$, $F(c\vec{u}) = cF(\vec{u})$.\\
 \begin{itemize}
 \item [i)] If $c = 0$ or $\vec{u} = \vec{0}_U$, this is obvious from $F(\vec{0}_U) = \vec{0}_V$.\\
 \item [ii)] If $c > 0$ and $\vec{u} \neq \vec{0}_U$, $F(c\vec{u})=\|c\vec{u}\| \vec{S}(c\vec{u}/\|c\vec{u}\|) = c\|\vec{u}\| \vec{S}(c\vec{u}/{c\|\vec{u}\|}) = c\|\vec{u}\| \vec{S}(\vec{u}/{\|\vec{u}\|}) = cF(\vec{u})$.\\ 
 \item [iii)] If $c < 0$, $F(c\vec{u})=-F(-c\vec{u})$ by a) above. But $-c > 0$ so $F(-c\vec{u})=-cF(\vec{u})$ by ii). Then $F(c\vec{u})=-(-c)F(\vec{u})=cF(\vec{u})$.\\
 \end{itemize}
 c) $F(\vec{u}_1+\vec{u}_2)=F(\vec{u}_1)+F(\vec{u}_2)$ for all $\vec{u}_1$, $\vec{u}_2 \in U$.\\
\begin{itemize}
\item [i)] If $\vec{u}_1 = \vec{0}_U$, $F(\vec{u}_1+\vec{u}_2)= F(\vec{u}_2)= \vec{0}_V+F(\vec{u}_2)= F(\vec{u}_1)+F(\vec{u}_2)$.\\
\item [ii)] If $\vec{u}_1 \neq \vec{0}_U$ and $\vec{u}_2=c\vec{u}_1$ then $F(\vec{u}_1+\vec{u}_2)=F((1+c)\vec{u}_1)=(1+c)F(\vec{u}_1)=F(\vec{u}_1)+cF(\vec{u}_1) =F(\vec{u}_1)+F(c\vec{u}_1)=F(\vec{u}_1)+F(\vec{u}_2)$.\\
\item [iii)] If $\{\vec{u}_1,\vec{u}_2\}$ is linearly independent, let $\vec{u}_3=-\vec{u}_1-\vec{u}_2$. We want to prove $F(-\vec{u}_3)=F(\vec{u}_1)+F(\vec{u}_2)$, or $-F(\vec{u}_3)=F(\vec{u}_1)+F(\vec{u}_2)$, or
  \begin{equation*}
F(\vec{u}_1)+F(\vec{u}_2)+F(\vec{u}_3)=\vec{0}_V. \qquad (13.2.24)
 \end{equation*}
\end{itemize}
 \begin{figure}[htbt]
\includegraphics[width=12cm]{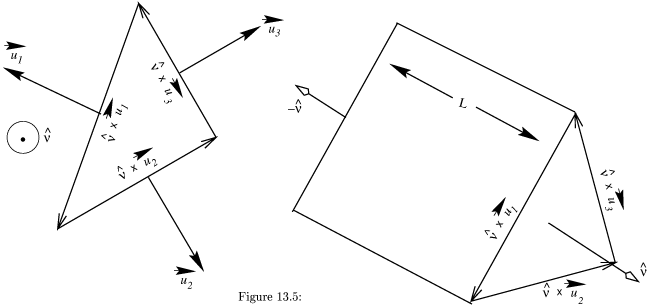}
\end{figure}
To prove $(13.2.24)$ note that since $\vec{u}_1$, $\vec{u}_2$ are linearly independent, we can define the unit vector $\hat{\nu}=(\vec{u}_1 \times \vec{u}_2)/{\|\vec{u}_1 \times \vec{u}_2\|}$.  We place the plane of this paper so that it contains $\vec{u}_1$ and $\vec{u}_2$, and $\hat{\nu}$ points out of the paper. The vectors $\vec{u}_1$, $\vec{u}_2$, $\vec{u}_3$ form the three sides of a nondegenerate triangle in the plane of the paper. $\hat{\nu}\times \vec{u}_i$ is obtained by rotating $\vec{u}_i$ $90^{\circ}$ counterclockwise. If we rotate the triangle with sides  $\vec{u}_1$, $\vec{u}_2$, $\vec{u}_3$ $90^{\circ}$ counterclockwise, we obtain a triangle with sides $\hat{\nu}\times \vec{u}_1$, $\hat{\nu}\times \vec{u}_2$, $\hat{\nu}\times \vec{u}_3$. The length of side $\hat{\nu}\times \vec{u}_i$ is $\|\hat{\nu}\times \vec{u}_i\|=\|\vec{u}_i\|$, and $\vec{u}_i$ is perpendicular to that side and points out of the triangle. Let $K'$ be the right cylinder whose base is the triangle with sides $\hat{\nu}\times \vec{u}_i$, and whose generators perpendicular to the base have length $L$. The base and top of the cylinder have area $A = \|\vec{u}_1 \times \vec{u}_2\|/2$ and their unit outward normals are $\vec{\nu}$ and $-\vec{\nu}$. The three rectangular faces of $K'$ have areas $L\|\vec{u}_i\|$ and unit outward normals $\vec{u}_{(i)}/\|\vec{u}_{(i)}\|$ Applying $(13.2.21)$ to this $K'$ gives
  \begin{equation*}
A\vec{S}(\hat{\nu})+ A\vec{S}(-\hat{\nu})+ \sum_{i=1}^{3} L\|\vec{u}_i\|\vec{S}(\vec{u}_i/\|\vec{u}_i\|)=\vec{0}_V.
 \end{equation*}
 But $\vec{S}(\hat{\nu})=-\vec{S}(-\hat{\nu})$ so dividing by $L$ and using $(13.2.22)$ gives $(13.2.23)$ [correction $(13.2.24)$].\\
 \newline
{\small {\bfseries Corollary $\mathbf{13.2.44}$}} (to Lemma $13.2.30.$) Under the hypotheses of lemma $13.2.30$, there is a unique $\overleftrightarrow{S}$ \ldots such that for all $\hat{n} \in N_U$
 \begin{equation*}
\vec{S}(\hat{n})=\hat{n}.\overleftrightarrow{S}. \qquad (13.2.25)
 \end{equation*}
}
\end{quote}
So, Backus uses two lemmas to prove the Cauchy's theorem. The first lemma $13.2.29$ leads to the fundamental integral equation $(13.2.15)$ for traction vectors at the exact point $\vec{r}_0$, that has some important enhancements as compared with other works. In the second lemma $13.2.30$, he tries to prove the existence of stress tensor based on equation $(13.2.15)$.

In the second lemma there is a process similar to the process in the Noll's proof in \cite{Tr-Gur} to prove the properties of the linear transformation for traction vectors that we have discussed it in the previous sections. But here this proof is on a different base from the Noll's proof. We saw that the Noll's proof was based on the infinitesimal volume and where the element's lengths approach zero \cite{Tr-Gur,Tr-Gen,Liu}. But here, Backus applies the Noll's equation $(13.2.22)$ to the obtained integral equation $(13.2.15)$ that is exactly valid for any volume of the mass element. Therefore, all of the relevant challenges to this step are removed in the Backus proof. 

A challenge is related to part (a) in the proof of lemma $13.2.30$, where in the equation:
  \begin{equation*}
L^2\vec{S}(\hat{n})+L^2\vec{S}(-\hat{n})+ \epsilon L (\vec{S}(\hat{n}_1)+\vec{S}(-\hat{n}_1)+\vec{S}(\hat{n}_2)+\vec{S}(-\hat{n}_2))=\vec{0}_V.
 \end{equation*}
the expression \emph{``Hold $L$ fixed and let $\epsilon \to 0$. Then divide by $L^2$ and $(13.2.23)$ is the result''}, may be interpreted as the result is valid only for a thin flat rectangular box (i.e., infinitesimal volume). But if we replace this expression by:
 
``Hold $L$ fixed and let $\epsilon$ change, it is not necessary that $\epsilon$ be a small value. Since the first two terms are independent of $\epsilon$, we must have $\vec{S}(\hat{n})+\vec{S}(-\hat{n})=\vec{0}_V$. So, $(13.2.23)$ is the result.''
 
This implies that the important equation $\vec{S}(-\hat{n})=-\vec{S}(\hat{n})$ is independent of the volume of mass element. Therefore, the challenges related to the infinitesimal volume are removed. Then, in parts (b) and (c), Backus proves exactly the essential properties of a linear transformation in vector space for $\vec{S}(\hat{n})$. Since a linear transformation in vector space can be shown by a second order tensor, the Backus's proof of the existence of stress tensor is completed. Also, in order to derive the differential equation of conservation of linear momentum, Backus uses the divergence theorem. 
\section{Conclusion}
In this article, we studied the tetrahedron arguments and the proofs of the existence of stress tensor in the literature. First, we showed the birth, importance and location of the tetrahedron argument and the existence of stress tensor in the foundation of continuum mechanics. By representation of the formal tetrahedron argument in detail, that is presented in many books, we extracted some fundamental challenges and discussed their importance. These conceptual challenges are related to the result of applying the conservation of linear momentum to any mass element in continuum media, the order of magnitude of the surface and volume terms in the integral equation of conservation of linear momentum, the definition of traction vectors on the surfaces that pass through the same point, the limited and approximate processes in the derivation of stress tensor, and some others. Then, in a comprehensive review of a large number of the relevant books during about two centuries from 1823 until now, we presented the different versions of tetrahedron argument and the proofs of the existence of stress tensor, and in each of them the challenges and the improvements are discussed. They can be classified in two general approaches. In the first approach, that is followed in most texts, the traction vectors are not defined on the surfaces that pass through the same point, but in a limited and approximate process when the volume of the mass element goes to zero, the traction vectors on the surfaces of the mass element are regarded as the traction vectors on the surfaces that pass through the same point. In the second approach, that is followed in a few books, the traction vectors are exactly defined at the same point on the different surfaces that pass through that point. Then in a limited and approximate process when the volume of the mass element goes to zero, a linear relation that leads to the existence of stress tensor, is obtained. By this approach some of the challenges are removed. We also presented and discussed the improved works of Hamel and Backus. Most of the challenges on the existence of stress tensor are removed in the unknown and original work of Backus. We presented the main parts of this proof and studied its improvements and challenges.
\bibliography{FirstPaper}{}

\begin{thebibliography}{100}

\bibitem{Antman}
S.~S. Antman.
\newblock {\em Nonlinear Problems of Elasticity}.
\newblock Springer, New York, 2005.

\bibitem{Arfken}
G.~Arfken.
\newblock {\em Mathematical Methods for Pysicists}.
\newblock Academic Press, Orlando, 1985.

\bibitem{Aris}
R.~Aris.
\newblock {\em Vectors, Tensors and the Basis Equations of Fluids Mechanics}.
\newblock Dover, 1989.

\bibitem{Atkin}
R.~J. Atkin and N.~Fox.
\newblock {\em An Introduction to the Theory of Elasticity}.
\newblock Courier Corporation, 2005.

\bibitem{Backus}
G.~Backus.
\newblock {\em Continuum Mechanics}.
\newblock Samizdat Press, USA, 1997.

\bibitem{Basar}
Y.~Başar and D.~Weichert.
\newblock {\em Nonlinear Continuum Mechanics of Solids: Fundamental
  Mathematical and Physical Concepts}.
\newblock Springer, Berlin New York, 2000.

\bibitem{Batchelor}
G.~K. Batchelor.
\newblock {\em An Introduction to Fluid Dynamics}.
\newblock Cambridge University Press, 2000.

\bibitem{Batra}
R.~C. Batra.
\newblock {\em Elements of Continuum Mechanics}.
\newblock American Institute of Aeronautics and Astronautics, Reston, Va, 2006.

\bibitem{Bechtel}
S.~E. Bechtel and R.~L. Lowe.
\newblock {\em Fundamentals of Continuum Mechanics: with Applications to
  Mechanical, Thermomechanical, and Smart Materials}.
\newblock Academic Press, 2014.

\bibitem{Biot}
M.~A. Biot.
\newblock {\em Mechanics of Incremental Deformations}.
\newblock Wiley, 1964.

\bibitem{Bonet}
J.~Bonet and R.~D. Wood.
\newblock {\em Nonlinear Continuum Mechanics for Finite Element Analysis}.
\newblock Cambridge University Press, Cambridge New York, 2008.

\bibitem{Borg}
S.~F. Borg.
\newblock {\em Matrix-Tensor Methods in Continuum Mechanics}.
\newblock D. Van Nostrand Company, 1966.

\bibitem{Brekhovskikh}
L.~M. Brekhovskikh and V.~Goncharov.
\newblock {\em Mechanics of Continua and Wave Dynamics}.
\newblock Springer-Verlag, Berlin New York, 1994.

\bibitem{Byskov}
E.~Byskov.
\newblock {\em Elementary Continuum Mechanics for Everyone: with Applications
  to Structural Mechanics}.
\newblock Springer, Dordrecht New York, 2013.

\bibitem{Calcote}
L.~R. Calcote.
\newblock {\em Introduction to Continuum Mechanics}.
\newblock D. Van Nostrand Company, 1968.

\bibitem{Capaldi}
F.~Capaldi.
\newblock {\em Continuum Mechanics: Constitutive Modeling of Structural and
  Biological Materials}.
\newblock Cambridge University Press, Cambridge, 2012.

\bibitem{Cau-1823}
A.~L. Cauchy.
\newblock Recherches sur l'{\'e}quilibre et le mouvement int{\'e}rieur des
  corps solides ou fluides, {\'e}lastiques ou non {\'e}lastiques.
\newblock {\em Bull Soc Filomat Paris 9–13}, 1823.

\bibitem{Cau-1827}
A.~L. Cauchy.
\newblock De la pression ou tension dans un corps solide (1822).
\newblock {\em Ex. de Math. 2, 42-56}, 1827.

\bibitem{Chandra}
D.~S. Chandrasekharaiah and L.~Debnath.
\newblock {\em Continuum Mechanics}.
\newblock Academic Press, 1994.

\bibitem{Chaves}
E.~Chaves.
\newblock {\em Notes on Continuum Mechanics}.
\newblock International Center for Numerical Methods in Engineering (CIMNE)
  Springer, Barcelona, Spain, 2013.

\bibitem{Ciarlet}
P.~Ciarlet.
\newblock {\em Mathematical Elasticity}.
\newblock North-Holland Sole distributors for the U.S.A. and Canada, Elsevier
  Science Pub. Co, 1988.

\bibitem{Isola}
F.~Dell’Isola, P.~Seppecher, and A.~Madeo.
\newblock Beyond euler-cauchy continua: The structure of contact actions in
  n-th gradient generalized continua: a generalization of the cauchy
  tetrahedron argument.
\newblock In {\em Variational Models and Methods in Solid and Fluid Mechanics},
  pages 17--106. Springer, 2011.

\bibitem{Dill}
E.~Dill.
\newblock {\em Continuum Mechanics: Elasticity, Plasticity, Viscoelasticity}.
\newblock CRC Press, Boca Raton, 2007.

\bibitem{Dimitrienko}
Y.~I. Dimitrienko.
\newblock {\em Nonlinear Continuum Mechanics and Large Inelastic Deformations}.
\newblock Springer, Dordrecht New York, 2011.

\bibitem{Epstein}
M.~Epstein.
\newblock {\em The Geometrical Language of Continuum Mechanics}.
\newblock Cambridge University Press, New York, 2010.

\bibitem{Eringen}
A.~C. Eringen.
\newblock {\em Mechanics of Continua}.
\newblock R.E. Krieger Pub. Co, Malabar, Fla, 1980.

\bibitem{Feynman}
R.~P. Feynman, R.~B. Leighton, and M.~L. Sands.
\newblock {\em The Feynman Lectures on Physics: Vol. 2: Ch. 31: Tensors}.
\newblock Addison-Wesley, 1965.

\bibitem{Flugge}
W.~Fl\"ugge.
\newblock {\em Tensor Analysis and Continuum Mechanics}.
\newblock Springer-Verlag, 1972.

\bibitem{Fosdick}
R.~L. Fosdick and E.~G. Virga.
\newblock A variational proof of the stress theorem of cauchy.
\newblock {\em Archive for Rational Mechanics and Analysis}, 105(2):95--103,
  1989.

\bibitem{Fung-1965}
Y.~C. Fung.
\newblock {\em Foundations of Solid Mechanics}.
\newblock Prentice Hall, 1965.

\bibitem{Fung-1969}
Y.~C. Fung.
\newblock {\em A First Course in Continuum Mechanics: for Physical and
  Biological Engineers and Scientists}.
\newblock Prentice Hall, Englewood Cliffs, N.J, 1994.

\bibitem{Godunov}
S.~Godunov and E.~I. Romenskii.
\newblock {\em Elements of Continuum Mechanics and Conservation Laws}.
\newblock Springer US, Boston, MA, 2003.

\bibitem{Gonzalez}
O.~Gonzalez and A.~M. Stuart.
\newblock {\em A First Course in Continuum Mechanics}.
\newblock Cambridge University Press, Cambridge, UK New York, 2008.

\bibitem{Graebel}
W.~Graebel.
\newblock {\em Advanced Fluid Mechanics}.
\newblock Academic Press, 2007.

\bibitem{Green}
A.~E. Green and W.~Zerna.
\newblock {\em Theoretical Elasticity}.
\newblock Courier Corporation, 1992.

\bibitem{Tr-Gur}
M.~E. Gurtin.
\newblock The linear theory of elasticity.
\newblock In {\em Linear Theories of Elasticity and Thermoelasticity}, pages
  1--295. Springer, 1973.

\bibitem{Gur-An}
M.~E. Gurtin.
\newblock {\em An Introduction to Continuum Mechanics}.
\newblock Academic Press, New York, 1981.

\bibitem{Gur-2010}
M.~E. Gurtin, E.~Fried, and L.~Anand.
\newblock {\em The Mechanics and Thermodynamics of Continua}.
\newblock Cambridge University Press, New York, 2010.

\bibitem{Gur-Mar}
M.~E. Gurtin and L.~C. Martins.
\newblock Cauchy's theorem in classical physics.
\newblock {\em Archive for Rational Mechanics and Analysis}, 60(4):305--324,
  1976.

\bibitem{Guyon}
E.~Guyon, J.~P. Hulin, L.~Petit, and C.~D. Mitescu.
\newblock {\em Physical Hydrodynamics}.
\newblock Oxford University Press, 2001.

\bibitem{Hamel}
G.~Hamel.
\newblock {\em Theoretische Mechanik}.
\newblock Springer, Berlin, 1949.

\bibitem{Han}
W.~Han-Chin.
\newblock {\em Continuum Mechanics and Plasticity}.
\newblock Chapman \& Hall/CRC, Boca Raton, 2005.

\bibitem{Haupt}
P.~Haupt.
\newblock {\em Continuum Mechanics and Theory of Materials}.
\newblock Springer Berlin Heidelberg, 2002.

\bibitem{Huilgol}
R.~R. Huilgol and N.~Phan-Thien.
\newblock {\em Fluid Mechanics of Viscoelasticity: General Principles,
  Constitutive Modelling, Analytical, and Numerical Techniques}.
\newblock Elsevier, 1997.

\bibitem{Hutter}
K.~Hutter and K.~J\"ohnk.
\newblock {\em Continuum Methods of Physical Modeling: Continuum Mechanics,
  Dimensional Analysis, Turbulence}.
\newblock Springer, Berlin New York, 2004.

\bibitem{Ilyushin}
A.~A. Ilyushin and V.~S. Lensky.
\newblock {\em Strength of Materials}.
\newblock Pergamon Press, 1967.

\bibitem{Irgens}
F.~Irgens.
\newblock {\em Continuum Mechanics}.
\newblock Springer, Berlin, 2008.

\bibitem{Jaunzemis}
W.~Jaunzemis.
\newblock {\em Continuum Mechanics}.
\newblock Macmillan, 1967.

\bibitem{Jog}
C.~S. Jog.
\newblock {\em Foundations and Applications of Mechanics, Vol 1: Continuum
  Mechanics}.
\newblock CRS Press Narosa Pub. House, Boca Raton, Fla. New Delhi, 2002.

\bibitem{Kiselev}
S.~P. Kiselev, E.~V. Vorozhtsov, and V.~M. Fomin.
\newblock {\em Foundations of Fluid Mechanics with Applications: Problem
  Solving Using Mathematica}.
\newblock Birkhauser Boston, 1999.

\bibitem{Kundu}
P.~K. Kundu, I.~M. Cohen, and D.~R. Dowling.
\newblock {\em Fluid Mchanics}.
\newblock Academic Press, Waltham, MA, 2012.

\bibitem{Lai}
W.~M. Lai, D.~Rubin, and E.~Krempl.
\newblock {\em Introduction to Continuum Mechanics}.
\newblock Butterworth-Heinemann/Elsevier, Amsterdam Boston, 2010.

\bibitem{Lautrup}
B.~Lautrup.
\newblock {\em Physics of Continuous Matter: Exotic and Everyday Phenomena in
  the Macroscopic World}.
\newblock Taylor \& Francis, Boca Raton, 2011.

\bibitem{Leal}
L.~G. Leal.
\newblock {\em Advanced Transport Phenomena: Fluid Mechanics and Convective
  Transport Processes}.
\newblock Cambridge University Press, New York, 2007.

\bibitem{Leigh}
D.~C. Leigh.
\newblock {\em Nonlinear Continuum Mechanics}.
\newblock New York., Mcgraw Hill, 1968.

\bibitem{Liu}
I.~S. Liu.
\newblock {\em Continuum Mechanics}.
\newblock Springer Berlin Heidelberg, 2002.

\bibitem{Love}
A.~E.~H. Love.
\newblock {\em A Treatise on the Mathematical Theory of Elasticity}.
\newblock Courier Corporation, 1944.

\bibitem{Lurie}
A.~I. Lurie.
\newblock {\em Theory of Elasticity}.
\newblock Springer, Berlin, 2005.

\bibitem{Malvern}
L.~E. Malvern.
\newblock {\em Introduction to the Mechanics of a Continuous Medium}.
\newblock Prentice Hall Inc., 1969.

\bibitem{Marsden}
J.~E. Marsden and T.~J.~R. Hughes.
\newblock {\em Mathematical Foundations of Elasticity}.
\newblock Courier Corporation, 1994.

\bibitem{Gur-Mar2}
L.~C. Martins.
\newblock On cauchy's theorem in classical physics: some counterexamples.
\newblock {\em Archive for Rational Mechanics and Analysis}, 60(4):325--328,
  1976.

\bibitem{Mase}
G.~T. Mase and G.~E. Mase.
\newblock {\em Continuum Mechanics for Engineers}.
\newblock CRC Press, Boca Raton, Fla, 1999.

\bibitem{Maugin}
G.~A. Maugin.
\newblock {\em Continuum Mechanics Through the Eighteenth and Nineteenth
  Centuries: Historical Perspectives from John Bernoulli (1727) to Ernst
  Hellinger (1914)}.
\newblock Springer, Cham, 2014.

\bibitem{Muskh}
N.~I. Muskhelishvili.
\newblock {\em Some Basic Problems of the Mathematical Theory of Elasticity:
  Fundamental Equations Plane Theory of Elasticity Torsion and Bending}.
\newblock Springer Netherlands Imprint Springer, Dordrecht, 1977.

\bibitem{Nair}
S.~Nair.
\newblock {\em Introduction to Continuum Mechanics}.
\newblock Cambridge University Press, New York, 2009.

\bibitem{Narasimhan}
M.~Narasimhan.
\newblock {\em Principles of Continuum Mechanics}.
\newblock Wiley, New York, 1993.

\bibitem{Noll-Light}
W.~Noll.
\newblock The foundations of classical mechanics in the light of recent
  advances in continuum mechanics.
\newblock {\em Studies in Logic and the Foundations of Mathematics},
  27:266--281, 1959.

\bibitem{Oden}
J.~T. Oden.
\newblock {\em An Introduction to Mathematical Modeling: A Course in
  Mechanics}, volume~1.
\newblock John Wiley \& Sons, 2011.

\bibitem{Prandtl}
H.~Oertel.
\newblock {\em Prandtl's Essentials of Fluid Mechanics}.
\newblock Springer, New York, 2004.

\bibitem{Ogden}
R.~W. Ogden.
\newblock {\em Non-linear Elastic Deformations}.
\newblock Courier Corporation, 1997.

\bibitem{Planck}
M.~Planck.
\newblock {\em The Mechanics of Deformable Bodies: being volume II of
  "Introduction to theoretical physics"}, volume~2.
\newblock Macmillan and Co., limited, 1932.

\bibitem{Reddy}
J.~N. Reddy.
\newblock {\em Principles of Continuum Mechanics: A Study of Conservation
  Principles with Applications}.
\newblock Cambridge University Press, New York, 2010.

\bibitem{Rivlin}
R.~S. Rivlin.
\newblock {\em Non Linear Continuum Theories in Mechanics and Physics and Their
  Applications}.
\newblock Springer, 1970.

\bibitem{Rudnicki}
J.~Rudnicki.
\newblock {\em Fundamentals of Continuum Mechanics}.
\newblock John Wiley \& Sons Inc, Chichester, West Sussex, United Kingdom,
  2015.

\bibitem{Salencon}
J.~Salençon.
\newblock {\em Handbook of Continuum Mechanics: General Concepts
  Thermoelasticity}.
\newblock Springer Berlin Heidelberg, 2001.

\bibitem{Schuricht}
F.~Schuricht.
\newblock A new mathematical foundation for contact interactions in continuum
  physics.
\newblock {\em Archive for Rational Mechanics and Analysis}, 184(3):495--551,
  2007.

\bibitem{Sedov-In}
L.~I. Sedov.
\newblock {\em Introduction to the Mechanics of a Continuous Medium}.
\newblock Addison-Wesley Reading, Mass., 1965.

\bibitem{Sedov-Cour}
L.~I. Sedov.
\newblock {\em A Course in Continuum Mechanics}.
\newblock Wolters-Noordhoff, Groningen, 1971.

\bibitem{Serrin}
J.~Serrin.
\newblock Mathematical principles of classical fluid mechanics.
\newblock In {\em Encyclopedia of Physics / Handbuch der Physik}, pages
  125--263. Springer Berlin Heidelberg, 1959.

\bibitem{Shames}
I.~H. Shames and F.~A. Cozzarelli.
\newblock {\em Elastic and Inelastic Stress Analysis}.
\newblock CRC Press, 1997.

\bibitem{Silhavy-84}
M.~{\v{S}}ilhavy.
\newblock The existence of the flux vector and the divergence theorem for
  general cauchy fluxes.
\newblock {\em Archive for Rational Mechanics and Analysis}, 90(3):195--212,
  1985.

\bibitem{Silhavy-90}
M.~{\v{S}}ilhav{\^y}.
\newblock On cauchy's stress theorem.
\newblock {\em Atti della Accademia Nazionale dei Lincei. Classe di Scienze
  Fisiche, Matematiche e Naturali. Rendiconti Lincei. Matematica e
  Applicazioni}, 1(3):259--263, 1990.

\bibitem{Silhavy-91}
M.~{\v{S}}ilhav{\`y}.
\newblock Cauchy's stress theorem and tensor fields with divergences in l p.
\newblock {\em Archive for Rational Mechanics and Analysis}, 116(3):223--255,
  1991.

\bibitem{Silhavy-08}
M.~{\v{S}}ilhav{\`y}.
\newblock Cauchy’s stress theorem for stresses represented by measures.
\newblock {\em Continuum Mechanics and Thermodynamics}, 20(2):75--96, 2008.

\bibitem{Slawinski}
M.~A. Slawinski.
\newblock {\em Waves and Rays in Elastic Continua}.
\newblock World Scientific, 2010.

\bibitem{Smith}
D.~Smith.
\newblock {\em An Introduction to Continuum Mechanics - after Truesdell and
  Noll}.
\newblock Springer Netherlands, Dordrecht, 1993.

\bibitem{Soko}
I.~S. Sokolnikoff.
\newblock {\em Mathematical Tehory of Elasticity}.
\newblock Tata Mcgraw Hill Publishing Company Ltd., Bombay, 1946.

\bibitem{Sommerfeld}
A.~Sommerfeld.
\newblock {\em Mechanics of Deformable Bodies: Lectures on Theoretical
  Physics}, volume~2.
\newblock Academic Press, New York, 1950.

\bibitem{Spencer}
A.~J.~M. Spencer.
\newblock {\em Continuum Mechanics}.
\newblock Dover Publications, Mineola, N.Y, 2004.

\bibitem{Stokes}
G.~G. Stokes.
\newblock On the theories of the internal friction of fluids in motion, and of
  the equilibrium and motion of elastic solid.
\newblock {\em Transactions of the Cambridge Philosophical Society}, Vol.
  8:287--319, 1945.

\bibitem{Talpaert}
Y.~Talpaert.
\newblock {\em Tensor Analysis and Continuum Mechanics}.
\newblock Springer Netherlands, 2002.

\bibitem{Temam}
R.~Temam and A.~Miranville.
\newblock {\em Mathematical Modeling in Continuum Mechanics}.
\newblock Cambridge University Press, 2005.

\bibitem{Timo-His}
S.~Timoshenko.
\newblock {\em History of Strength of Materials: with a Brief Account of the
  History of Theory of Elasticity and Theory of Structures}.
\newblock Courier Corporation, 1953.

\bibitem{Timo-Theo}
S.~Timoshenko and J.~N. Goodier.
\newblock {\em Theory of Elasticity}.
\newblock Mcgraw Hill, 1951.

\bibitem{Todhunter}
I.~Todhunter and K.~Pearson.
\newblock {\em A History of the Theory of Elasticity and of the Strength of
  Materials}.
\newblock Cambridge University Press, 1886.

\bibitem{Tr-Ess}
C.~Truesdell.
\newblock {\em Essays in the History of Mechanics}.
\newblock Springer-Verlag New York Inc, 1968.

\bibitem{Tr-Ther}
C.~Truesdell.
\newblock {\em The Tragicomedy of Classical Thermodynamics}.
\newblock Springer, 1971.

\bibitem{Tr-Gen}
C.~Truesdell.
\newblock {\em A First Course in Rational Continuum Mechanics, Vol I, General
  Concepts}.
\newblock Academic Press, Boston, 1991.

\bibitem{Tr-Cla}
C.~Truesdell and R.~Toupin.
\newblock The classical field theories.
\newblock In {\em Principles of Classical Mechanics and Field Theory /
  Prinzipien der Klassischen Mechanik und Feldtheorie}, pages 226--858.
  Springer, 1960.

\bibitem{Wang}
C.~C. Wang.
\newblock {\em Mathematical Principles of Mechanics and Electromagnetism: Part
  A: Analytical and Continuum Mechanics}.
\newblock Springer US, Boston, MA, 1979.

\bibitem{Wegner}
J.~Wegner and J.~B. Haddow.
\newblock {\em Elements of Continuum Mechanics and Thermodynamics}.
\newblock Cambridge University Press, New York, 2009.

\end{thebibliography}
\bibliographystyle{plain}
\end{document}